\documentclass[secnumarabic,twocolumn, graphics,floatfix,tightenlines,aps,prb,superscriptaddress,longbibliography]{revtex4-1}

\bibpunct{[}{]}{;}{n}{}{}

 \pdfoutput=1

\usepackage{epsfig}
\usepackage{amsmath,amssymb,amsfonts}
\usepackage{mathrsfs}
\usepackage{bbm}
\usepackage{slashed}
\usepackage{graphicx}
\usepackage{verbatim}
\usepackage[usenames]{color}
\usepackage{mathtools}

\usepackage{tabularx}

\usepackage{stackengine}
\usepackage{hyperref}
\usepackage{bookmark}

\usepackage{bibunits}

\usepackage{blindtext}

\newcommand\underrel[2]{\mathrel{\mathop{#2}\limits_{#1}}}

\newcommand{\scr}[1]{\mathcal{#1}}

\newcommand{\p}{\partial}

\newcommand{\LCm}{{\scriptscriptstyle -}} %LC supersripts
\newcommand{\LCp}{{\scriptscriptstyle +}}

\newcommand{\LCpm}{{\scriptscriptstyle \pm}}

\newcommand{\ve}[1]{{\bf{#1}}}

\newcommand{\av}[1]{\langle #1 \rangle}

\newcommand{\id}{\text{d}}

\newcommand{\G}{\scr{G}}

\newcommand{\QAx}{\mathrm{{\bf Q}}_1}
\newcommand{\QAy}{\mathrm{{\bf Q}}_2}
\newcommand{\QDx}{\overline{\mathrm{{\bf Q}}}_1}
\newcommand{\QDy}{\overline{\mathrm{{\bf Q}}}_2}

\newcommand{\ov}[1]{\overline{#1}}
\newcommand{\ti}[1]{\tilde{#1}}

\newcommand{\sectA}{\text{A}}
\newcommand{\sectB}{\text{B}}

\renewcommand\labelenumi{(\roman{enumi})}
\renewcommand\theenumi\labelenumi

\begin{document}
	\title{Nematic single-component superconductivity and loop-current order from pair-density wave instability}
	\author{Jonatan W\aa rdh}
	\email[]{jonatan.wardh@gmail.com}
	\affiliation{Department of Physics, University of Gothenburg,
		SE-41296 Gothenburg, Sweden}
	%
%	\author{Brian M. Andersen}
%	\email[]{bma@nbi.ku.dk}
%	\affiliation{Niels Bohr Institute, University of Copenhagen, Juliane Maries Vej 30, DK-2100 Copenhagen, Denmark}
	%
	\author{Mats Granath}
	\email[]{mats.granath@physics.gu.se}
	\affiliation{Department of Physics, University of Gothenburg,
		SE-41296 Gothenburg, Sweden}
	
\begin{abstract}
% BARA 600 characters including spaces! for PRL
%
We investigate the nematic and loop-current type orders that may arise as vestigial precursor phases in a model with an underlying pair-density wave (PDW) instability. We discuss how such a vestigial phase gives rise to a highly anisotropic stiffness for a coexisting single-component superconductor with low intrinsic stiffness, as is the case for the underdoped cuprate superconductors. Next, focusing on a regime with a mean-field PDW ground state with loop-current and nematic $xy$ (B$_{2g}$) order, we find a preemptive transition into a low and high-temperature vestigial phase with loop-current and nematic order corresponding to $xy$ (B$_{2g}$) and $x^2-y^2$ (B$_{1g}$) symmetry respectively. Near the transition between the two phases, a state of soft nematic order emerges for which we expect that the nematic director is readily pinned away from the high-symmetry directions in the presence of an external field. Results are discussed in relation to findings in the cuprates, especially to the recently inferred highly anisotropic superconducting fluctuations [W{\aa}rdh {\em et al.}, ``Colossal transverse magnetoresistance due to nematic superconducting phase fluctuations in a copper oxide'', arXiv:2203.06769], giving additional evidence for an underlying ubiquitous PDW instability in these materials. 

\end{abstract}
\maketitle

\section{Introduction}
One major challenge in the study of cuprate high-temperature superconductors is to unravel the intricate interplay of "intertwined" electronics orders \cite{fradkin2015colloquium}, and their relation to the pseudogap. Spin and charge orders have been shown to be ubiquitous phenomena in these compounds  \cite{tranquada1995evidence,ghiringhelli2012long,chang2012direct,le2014inelastic},
%RevModPhys.75.1201,arpaia2019dynamical
 as well as nematic order \cite{keimer2015quantum,lawler2010intra,fujita2014simultaneous,zheng2017study,doi:10.1126/science.abc8372}.
Another pertinent electronic order is the spatially modulated superconducting state, known as a pair-density wave (PDW)\cite{agterberg2020physics,himeda2002stripe}, which is conceptually related to the Fulde-Ferrell-Larkin-Ovchinnikov\cite{fulde1964superconductivity,larkin1965inhomogeneous} (FFLO) type states.  The PDW came to prominence in the cuprate context to explain the anomalous suppression of superconductivity at $1/8$ doping in the striped superconductor La$_{2-x}$Ba$_{x}$CuO$_4$ \cite{berg2007dynamical,berg2009theory,moodenbaugh1988superconducting,tranquada1995evidence,li2007two}. More recently, to explain the apparent residual superconductivity in the pseudogap, in the form of a prevalent diamagnetic response \cite{li2010diamagnetism}, together with the omnipresent charge-density wave (CDW), PDW order has also been suggested as the "mother state" of the pseudogap itself\cite{chakravarty2001hidden,lee2014amperean}. Related to this, PDW has been discussed in the context of Fermi-arcs \cite{baruch2008spectral}, and the anomalous quantum oscillations at large magnetic fields\cite{zelli2011mixed,norman2018quantum,PhysRevResearch.3.023199}. More direct signatures have been reported based on scanning tunneling spectroscopy\cite{hamidian2016detection,edkins2018magnetic}. 
Furthermore, numerous evidence points towards a time-reversal breaking intra-unit cell magnetic order present in the pseudogap phase \cite{kaminski2002spontaneous,fauque2006magnetic,li2008unusual,li2011magnetic,sidis2013evidence,mangin2014characterization}. This has spurred the suggestion of various kinds of magnetoelectric(ME) orders, specifically so-called loop-current orders  \cite{varma2000proposal,simon2002detection,varma2006theory,orenstein2011optical,yakovenko2015tilted}, which breaks time-reversal symmetry and parity, but preserves their product.
% yakovenko2015tilted, tilted loop cuurent to account for the Kerr- effect \cite{xia2008polar} Review by kapitulik \cite{kapitulnik2009polar}

Another recent theme in the physics of strongly correlated materials and the cuprates is that of vestigial orders, which refers to the emergence of a secondary order parameter that breaks a subgroup of symmetries of a multicomponent order parameter at a critical temperature that may surpass that of the underlying order. Such discrete broken symmetry has been discussed both in the context of nematic order\cite{fernandes2012preemptive,fernandes2014drives,fernandes2019intertwined} and broken time reversal symmetry\cite{PhysRevB.81.134522,PhysRevB.88.220511,PhysRevB.89.104509,https://doi.org/10.48550/arxiv.2102.06158}, as well as partially broken continuous symmetry phases of multicomponent superconductors\cite{babaev2004superconductor}. 

Vestigial order is natural to appeal to as a source for intra-cell order when the multicomponent order is related to the point group of the lattice and have been studied as a source of nematicity, with  evidence in  iron-based, topological, and cuprate superconductors\cite{fernandes2019intertwined,shibauchi2020exotic,fernandes2022iron,cho2020z3,mukhopadhyay2019evidence}. In hole doped Ba$_{1-x}$K$_x$Fe$_2$As$_2$ a recent study shows evidence for a state with incoherent pairing but broken time reversal symmetry consistent with a vestigial state of a multiband superconductor\cite{grinenko2021state}. In the cuprates, vestigial-nematic order has been suggested to possibly arise both from spin and charge order\cite{nie2014quenched,PhysRevB.96.085142}, as well as PDW\cite{agterberg2015emergent}. It has also been shown that loop-current orders can arise as a vestigial order that preempts a magnetoelectric PDW state (ME-PDW)\cite{agterberg2015emergent}.
\begin{figure}[h!]
	\centering
	\includegraphics[width=0.4\textwidth]{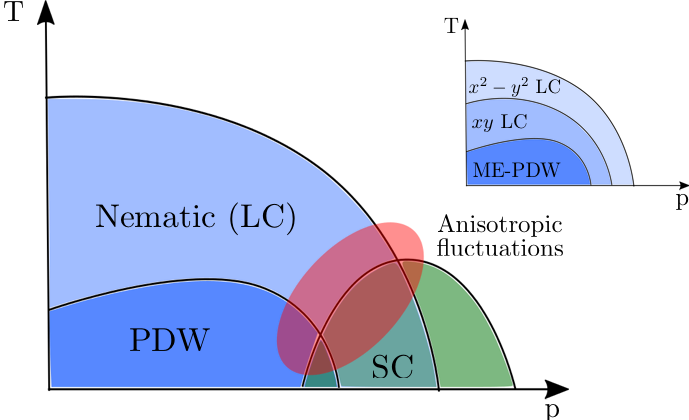} 
	\caption{\label{fig_pahase_diagram} \footnotesize Possible interpretation of cuprate phase diagram with a PDW as the mean-field pseudogap state setting up vestigial nematic, as well as loop-current(LC) order. Near and above $T_c$ in the underdoped part of the superconducting(SC) dome, this is consistent with the anisotropic superconducting fluctuations seen in LSCO\cite{wu2017spontaneous,wardh2019inprep}, due to the closeness of a PDW instability and presence of nematic order. The inset shows the possibility of splitting the vestigial nematic phase into a low and high temperature phase of $xy$(B$_{2g}$) and $x^2-y^2$(B$_{1g}$) nematic order respectively, due to an underlying magnetoelectric PDW (ME-PDW).
	}
\end{figure}
\begin{figure*}
	\begin{tabular}{cc}
		\begin{minipage}{0.7\textwidth}
			\includegraphics[width=0.9\textwidth]{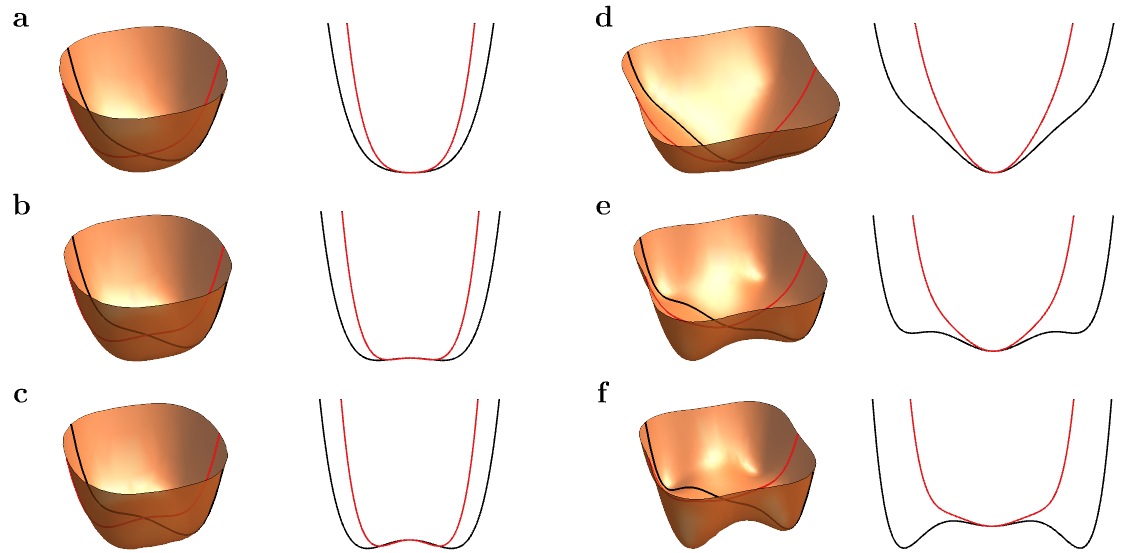} 
		\end{minipage}
		&
		\begin{minipage}{0.3\textwidth}
			\includegraphics[width=0.8\textwidth]{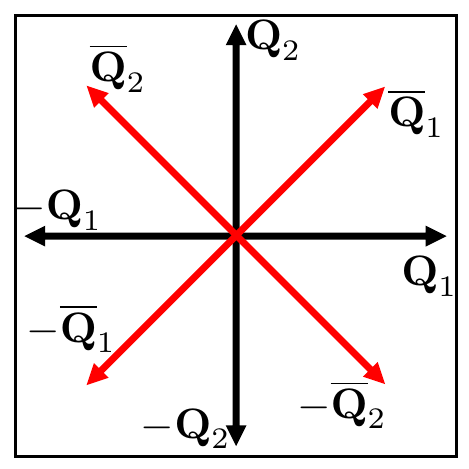} 
		\end{minipage}
		\\
	\end{tabular}
	\begin{minipage}{1.0\textwidth}
		\caption{\label{Figure_pot} The dispersion \eqref{eq_dispersion} shown for various $a$ and $c$, and $d>0,e=, b>0$. {\bf Left} Dispersion along the crystal axis, $x$, (black), and along the diagonal (red). 
		%The condition $b>0$ ensures that the minima will lie along axis (black), as supposed to the diagonal directions (red). 
		{\bf a,b,c} Successive decrease of $a$ for $c>0$. A finite $q$ develops continuously from zero. {\bf d,e,f} Successive decrease of $a$ for $c<0$. For big enough $a$, there is only a stable state at $q=0$ {\bf(d)}, by decreasing $a$ a metastable $q \neq 0$ state develops {\bf(e)}, which eventually becomes the new stable state {\bf(f)}. {\bf Right} The corresponding PDW ordering vectors.}
	\end{minipage}
\end{figure*}

In this paper we explore the occurrence and competition between various PDW-vestigial phases, motivated both by an explicit model for stabilizing PDW order based on pair-hopping interactions\cite{waardh2017effective,waardh2018suppression} (see also \cite{wu2022pair}), and by recent experiments on strongly nematic phase fluctuations in a cuprate superconductor\cite{wardh2019inprep}. 
First, we show how the relation between PDW and homogeneous superconductivity naturally generates an anisotropic superconducting state in a vestigial nematic PDW phase. In turn, this anisotropy can become strongly enhanced due to low  phase-stiffness\cite{emery1995importance,uemura1989universal}, which in itself may also be related to a proximate PDW instability\cite{waardh2017effective,waardh2018suppression}. Transport measurements in La$_{2-x}$Sr$_{x}$CuO$_4$ (LSCO) have shown evidence of an electronic nematic order\cite{wu2017spontaneous}. These measurements indicate highly anisotropic superconducting fluctuations near the underdoped critical point, with only a small anisotropy of the normal electrons\cite{wardh2019inprep}. This is consistent with the collective dynamics of the superconductor being highly susceptible to nematic order, along the lines presented in the present paper. A caricature of a phase diagram based on such a scenario, with interplay between superconductivity and vestigial PDW order, is presented in Figure \ref{fig_pahase_diagram}.

In tetragonal symmetry, loop-current (LC) order is associated with a vector $\vec{l}=(l_x,l_y)$, transforming in the E$_u$ representation. In the second part of the paper we will explore a scenario with an underlying ME-PDW state with LC order, which is invariant under reflection in the crystallographic diagonal $xy$ ($l_x=l_y$), with subleading $xy$ (or B$_{2g}$) nematic order $l_x l_y$. This phase is naturally preempted by a phase without the long-range PDW order but with vestigial LC and nematic order\cite{agterberg2015emergent}, which we refer to as an $xy$ LC phase.
% (in terms of broken symmetries it corresponds to the so-called $\Theta_{2}$ state \cite{simon2002detection,varma2006theory})
%
We show that this preemptive transition can be split further into a low and and high-temperature phase. The low-temperature phase coincides with the $xy$ LC phase, while the high-temperature phase breaks the tetragonal symmetry differently, by developing LC order which is invariant under reflection in the crystallographic axis $l_x\neq 0, l_y=0$, i.e.  $x^2-y^2$ (or B$_{1g}$) nematic order $l_x^2 - l_y^2$. We refer to this as an $x^2-y^2$ LC phase. Besides giving a richer phase diagram with both B$_{1g}$ and B$_{2g}$ symmetric orders, arising from the same underlying ME-PDW state, we find near the first-order transition between the $x^2-y^2$ and $xy$ LC phase a state with approximate rotational symmetry in $\vec{l}$. This yields a very soft nematic order with is highly susceptible to external fields that may pin the nematic director away from the high symmetry directions.

\subsection{Outline}

This paper is composed of two main results parts and is outlined as follows. In Section \ref{section_model} the considered model is discussed. This model is based on the phenomenology of an instability to a PDW state developed in\cite{waardh2017effective,waardh2018suppression}, but shares features with other discussed models for a PDW state\cite{lee2014amperean,agterberg2015checkerboard,https://doi.org/10.48550/arxiv.2110.13138,setty2022exact,wu2022pair}. In the first results part, Section \ref{section_an_sc}, the model is decomposed into possible vestigial order parameters which is then used to develop an effective model of a uniform, but anisotropic, superconducting state, Eqn.~\ref{eq_effective_superconductive_action}. In Section \ref{section_super_fluct} we discuss how the proximity to a PDW instability, with concomitant vestigial nematic order, can give rise to a very large stiffness anisotropy of the superconductor. 

In the second results part, Section \ref{section_enhanced_sym}, we explore the possible vestigial phases to a PDW state, implicitly assumed in Section \ref{section_an_sc}. Here we focus on a parameter regime where the $xy$ ME-PDW is stable, exploring its potential vestigial phases.

\section{Model \label{section_model}}
A PDW state denotes a state where paired electrons have finite momenta $\Delta_{\ve{Q}} \sim \av{c_{\downarrow, \ve{k}}c_{\uparrow, -\ve{k}+\ve{Q}}}$. We consider the situation when a PDW state (with a spatially modulated superconducting order) is near degenerate with a homogeneous superconducting state. The partition function takes the form $Z=\int \scr{D} \Delta \, e^{-S}$ with
%
%\begin{widetext}
\begin{equation}
\begin{split}
%\\
%
S&=\frac{1}{T}\int \limits_{\vec{x}} r_0|\Delta(\vec{x})|^2+\Delta^{*}(\vec{x})\varepsilon(\vec{D})\Delta(\vec{x}) \\
+ &\frac{1}{2T}\int \limits_{\substack{\vec{x}_1,\vec{x}_2 \\\vec{x}_3,\vec{x}_4}} U(\vec{x}_1,\vec{x}_2,\vec{x}_3,\vec{x}_4) \Delta(\vec{x}_1) \Delta^{*}(\vec{x}_2) \Delta(\vec{x}_3) \Delta^{*}(\vec{x}_4) \\
%
%L&=\int \limits_{k} (r+\varepsilon(k))|\Delta(k)|^2 + \int \limits_{k_1,k_2,k_3}\frac{u(k_1,k_2,k_3)}{2} \Delta(k_1) \Delta^{*}(k_2) \Delta(k_3) \Delta^{*}(k_1-k_2+k_3) \\
\end{split}
%\label{eq_full_partition}
\label{eq_action_sc}
\end{equation}
%\end{widetext}
%

% Philosophy: \beta = 1 
where $\vec{D}=-i \vec{\nabla}-2e \vec{A}$. The action $S$ is given by the most general (2D) Ginzburg-Landau expression, with interaction $U$, to fourth order in superconducting order, which respects $U(1)$ gauge symmetry, translational symmetry, and the point-group symmetry $D_{4\text{h}}$. 
In order to describe an instability towards PDW order the superconducting "dispersion" $\varepsilon(\vec{p})$ should develop minima at finite momenta. We consider a general dispersion to sixth order in in momenta $\vec{p}=(p_x,p_y)$
\begin{equation}
\varepsilon(\vec{p})=a p^2+b p_x^2p_y^2+c (p^2)^2+d (p^2)^3+e (p_x^2 p_y^4+ p_y^2 p_x^4)   \, ,
\label{eq_dispersion}
\end{equation}
to ensure stability we take $d>0,e>-4d$. This renormalized dispersion naturally occur in models with pair-hopping interactions\cite{waardh2017effective,waardh2018suppression}, but can also be considered a phenomenological model for coexisting zero- and finite-momentum superconductivity.   

The instability to PDW order can occur in two different ways. The first, as shown in Figure \ref{Figure_pot}a,b and c, is a continuous evolution of the pairing momenta from $p=0$ to finite $p$, parametrized by $a$ going from positive to negative for $c>0$. When $a=0$ the dispersion of the homogeneous superconductor becomes flat, constituting a Lifshitz-point, where the stiffness to fluctuations goes to zero.
The second, occurring when $c<0$ and shown in Figure \ref{Figure_pot}d,e and f, is a discontinuous jump in $p$ through the development of a distinct metastable state at finite momenta $p \neq 0$, which becomes stable when $a$ is decreased sufficiently.
The dispersion \eqref{eq_dispersion} allows local minima both along the axes and the diagonals. In general eight finite momentum vectors are allowed $\pm\ve{Q}_1,\pm\ve{Q}_2,\pm\overline{\ve{Q}}_1,\pm\overline{\ve{Q}}_2$ given by $\ve{Q}_1=Q(1,0)$, $\ve{Q}_2=Q(0,1)$, $\overline{\ve{Q}}_1=\overline{Q}(1,1)/\sqrt{2}$ and $\overline{\ve{Q}}_2=\overline{Q}(-1,1)/\sqrt{2}$. 

In order to analyze the fluctuations near the onset of these finite momentum orders we will go on re-expressing \eqref{eq_action_sc} by expanding in the various finite momentum superconducting order parameters. We will, however, leave the explicit form of the fourth order term $U$ in \eqref{eq_action_sc} unspecified and instead infer the expansion by considering all symmetry allowed terms. Before writing down the expression for the expanded action, we first discuss formally what terms are allowed. 

\subsection{Symmetry and order parameters}
We consider the tetragonal point-group symmetry $D_{4\text{h}}$ generated by $\{C_4, \sigma_{v},\sigma_{h}\}$, where $C_{4}$ is a four-fold rotation about the $z$-axis, $\sigma_{v}$ reflection in the $xz(yz)$ plane and $\sigma_{h}$ reflection in the $xy$ plane. The different momenta of the PDW order leads to eight different complex order parameters, and one ordinary homogeneous superconducting field $\Delta_0$. The latter is assumed to be a single-component complex field, transforming in a one-dimensional representation of the point group, e.g.\ a $d_{x^2-y^2}$ wave order. The set of order parameters, $\Gamma$, is divided into three sectors, \sectA, \sectB, and SC, $\Gamma = \Gamma_{\sectA} \oplus \Gamma_{\sectB } \oplus \Delta_0$. $\Gamma_{\sectA} =\{\Delta_{\ve{Q}_1},\Delta_{-\ve{Q}_1},\Delta_{\ve{Q}_2}, \Delta_{-\ve{Q}_2} \}$ and $\Gamma_{\sectB } =\{\Delta_{\ov{\ve{Q}}_1},\Delta_{-\ov{\ve{Q}}_1},\Delta_{\ov{\ve{Q}}_2}, \Delta_{-\ov{\ve{Q}}_2} \}$ contains the PDW fields, and do not transform into each other under $D_{4\text{h}}$, but their form is related by a $45^{\circ}$ twist. These are denoted by black and red in Figure \ref{Figure_pot}. Besides the point-group symmetry, the action will be invariant under $U(1)$ gauge symmetry, translational symmetry and time-reversal symmetry. Under these symmetries the order parameters transforms as $\Delta_{\ve{Q}}\xrightarrow[]{U(1)} \Delta_{\ve{Q}}e^{i \theta}$, $\Delta_{\ve{Q}}\xrightarrow[]{\text{T}} \Delta_{\ve{Q}}e^{i \ve{T}\cdot \ve{Q}}$, and $\Delta_{\ve{Q}}\xrightarrow[]{\scr{T}} \Delta^{*}_{-\ve{Q}}$.
% and $\Delta_{\ve{Q}}\xrightarrow[]{\scr{P}} \Delta_{-\ve{Q}}$. 
% parity is also a point group operation

\subsubsection{Composite order parameters}
\begin{table}
	\begin{tabular}{l | l}
	\hline			
	\hline
	Bilinears & Irrep.   \\
	\hline
	$|\Delta_0|^2$ & A$_{1g}$, $z^2$ \\ \hline
	$\psi_{\sectB }: |\Delta_{\QAx}|^2 + |\Delta_{-\QAx}|^2 + |\Delta_{\QAy}|^2 + |\Delta_{-\QAy}|^2$ & A$_{1g}$, $x^2\! \! \LCp y^2$ \\
	$N_{x^2 \! \! \LCm y^2}: |\Delta_{\QAx}|^2 + |\Delta_{-\QAx}|^2 -|\Delta_{\QAy}|^2 - |\Delta_{-\QAy}|^2$ & B$_{1g}$, $x^2 \! \! \LCm y^2$ \\
	$\vec{l}_{\sectA }: \Big[|\Delta_{\QAx}|^2 - |\Delta_{-\QAx}|^2, |\Delta_{\QAy}|^2 - |\Delta_{-\QAy}|^2 \Big]$
	& E$_u$, $(x,y)$ \\ \hline
	$\psi_{\sectA }: |\Delta_{\QDx}|^2 + |\Delta_{-\QDx}|^2 + |\Delta_{\QDy}|^2 + |\Delta_{-\QDy}|^2$ & A$_{1g}$, $x^2\! \! \LCp y^2$ \\
	$N_{xy}: |\Delta_{\QDx}|^2 + |\Delta_{-\QDx}|^2 -|\Delta_{\QDy}|^2 - |\Delta_{-\QDy}|^2$ & B$_{2g}$, $xy$ \\
	$\vec{l}_{\sectB }:$
	\begin{tabular}{c}
	$\Big[ \frac{|\Delta_{\QDx}|^2 - |\Delta_{-\QDx}|^2-|\Delta_{\QDy}|^2 + |\Delta_{-\QDy}|^2}{\sqrt{2}}, $ \\
	$\frac{|\Delta_{\QDx}|^2 - |\Delta_{-\QDx}|^2+|\Delta_{\QDy}|^2 - |\Delta_{-\QDy}|^2}{\sqrt{2}} \Big] $ \\
	%
	%$=\Big[ \frac{l_{\sectB  \ov{x}}-l_{\sectB  \ov{y}}}{\sqrt{2}}, \frac{l_{\sectB  \ov{x}}+l_{\sectB  \ov{y}}}{\sqrt{2}}\Big] $ \\
	\end{tabular}
	& E$_u$, $(x,y)$ \\
	\hline  
	\hline
\end{tabular}
	\caption{\label{table} Possible irreducible representations of the set of composite orders (bilinears)
		$\Gamma^{(2)}=\Gamma^{(2)}_{\sectA}\oplus\Gamma^{(2)}_{\sectB } \oplus |\Delta_0|^2$, where $\Gamma^{(2)}_{\sectA} =\{|\Delta_{\ve{Q}_1|^2},|\Delta_{-\ve{Q}_1}|^2,|\Delta_{\ve{Q}_2}|^2, |\Delta_{-\ve{Q}_2}|^2 \}$ and $\Gamma^{(2)}_{\sectB } =\{|\Delta_{\ov{\ve{Q}}_1}|^2,|\Delta_{-\ov{\ve{Q}}_1}|^2,|\Delta_{\ov{\ve{Q}}_2}|^2, |\Delta_{-\ov{\ve{Q}}_2}|^2 \}$. %urrent order for the diagonal sector using the diagonal basis vectors $\ov{x}=\frac{\hat{x}+\hat{y}}{\sqrt{2}}, \ov{y}=\frac{-\hat{x}+\hat{y}}{\sqrt{2}}$. %\com{Comment on last row}
		}
\end{table}

The action will be made up of all possible products of $\Gamma$, that transform trivially under the full symmetry group $U(1) \otimes T \otimes \scr{T} \otimes D_{4\text{h}}$. These will be second order, $\Gamma^{*} \otimes \Gamma$, and fourth order terms $\Gamma^{*} \otimes \Gamma \otimes \Gamma^{*} \otimes \Gamma$. (Terms including derivatives are discussed in Section \ref{section_derivative}.) The possible vestigial phases will be described by a set of order parameters, $\{ \phi_1,\phi_2... \}$, which are to second order in the primary fields $\phi \sim \Gamma^{*} \otimes \Gamma$, and transforming in non-trivial irreducible representations. We re-express the action in these composite order parameters, integrating out the PDW fields, $\Gamma$. The new action will thus be made up of products of these composite order parameters, that transform trivially under the full symmetry group. The breaking of symmetries and emergence of vestigial phases is then understood in the language of a Landau phase transition $L= a \phi^2 +b\phi^4$ where the order parameter $\phi$ develops a non-zero expectation value for $a<0$, thus breaking the corresponding symmetry of the system.

We are especially interested in composite orders that only break the point-group symmetries, i.e. non-superconducting intra-unit cell orders. There are 9 bilinears that transforms trivially under $U(1) \otimes T$, which we write as
%We will therefore directly restrict the space of $9 \times 9$ possible bilinears to $9$, which transforms trivially under $U(1) \otimes T$
% 
$\Gamma^{(2)}=\Gamma^{(2)}_{\sectA}\oplus\Gamma^{(2)}_{\sectB } \oplus |\Delta_0|^2$, where $\Gamma^{(2)}_{\sectA} =\{|\Delta_{\ve{Q}_1}|^2,|\Delta_{-\ve{Q}_1}|^2,|\Delta_{\ve{Q}_2}|^2, |\Delta_{-\ve{Q}_2}|^2 \}$ and $\Gamma^{(2)}_{\sectB } =\{|\Delta_{\ov{\ve{Q}}_1}|^2,|\Delta_{-\ov{\ve{Q}}_1}|^2,|\Delta_{\ov{\ve{Q}}_2}|^2, |\Delta_{-\ov{\ve{Q}}_2}|^2 \}$. Again, $\Gamma^{(2)}_{\sectA}$ and $\Gamma^{(2)}_{\sectB }$ do not mix under $D_{4 \text{h}}$ and can be decomposed further into their irreducible representations: $\Gamma^{(2)}_{\sectA}  = A_{1g} \oplus B_{1g} \oplus E_u$ and $\Gamma^{(2)}_{\sectB }  = A_{1g} \oplus B_{2g} \oplus E_u$, which are listed in Table \ref{table}.

%\subsubsection{Nematic and loop-current order and the ME-PDW state \label{section_nematic_loop}}
%
The decomposition into bilinears implies the existence of two nematic order parameters, transforming as B$_{1g}$ and B$_{2g}$, as well as two polar vector orders, transforming as E$_u$. The polar-vector order $l_i = |\Delta_{\ve{Q}_i}|^2 -  |\Delta_{-\ve{Q}_i}|^2$ is odd under parity and has the symmetry of a toroidal moment, which shares symmetry with the so-called loop-current (LC) order, and we will refer to it as such. 
%An LC order can classically be represented by $\ve{L} \propto  \! \!  \! \! \!  \! \! \!  \int \limits_{\text{unit cell}} \! \! \!  \! \! \! \id \ve{r} \, \ve{M} \times \ve{r}$, where $\ve{M} \propto \ve{r} \times \ve{p}$ is the magnetic moment\cite{shekhter2009considerations}.%   from moving charges
We will refer to an ME-PDW as a PDW with finite expectation value on LC order.

\subsubsection{Derivative terms \label{section_derivative}}
Derivative terms arise by forming products between $\vec{D}=(D_x,D_y)$ (transforming as E$_u$), and the bilinears $\Gamma^{(2)}$. For $\Delta_0$, transforming as A$_{1g}$, no linear derivative terms can arise to any order in $\Delta_0$. Mixing with PDW bilinears, transforming as $2 \text{A}_{1g} \oplus \text{B}_{1g} \oplus \text{B}_{2g} \oplus 2 \text{E}_u$, linear derivatives are allowed both to second and fourth order in fields. But, terms linear in derivatives implies an instability of the PDW momenta. Thus, expanding around stable local minima of \eqref{eq_dispersion}, will to first non-vanishing order generate second-order terms in derivative and fields.

We do have the possibility of including derivative terms that are to fourth order in fields. However, usually, these terms are irrelevant compared to derivative terms arising to second order in fields. We will assume this is still true for the PDW fields, for which these terms will be neglected. However, near the Lifshitz point, where the dispersion for $\Delta_0$ becomes flat, derivative terms acting on $\Delta_0$, occurring to fourth order in fields, will be important to include. To second order in derivatives and to fourth order in fields,
we can form products between an A$_{1g}$ derivative term and the A$_{1g}$ bilinears, or the B$_{1g}$(B$_{2g}$) derivative term with the B$_{1g}$(B$_{2g}$) bilinears. The first term contributes to the isotropic stiffness and is of no particular interest (it can be included in the overall renormalization), the second type of term will, however, generate an anisotropic stiffness in the presence of (vestigial) nematic order from the PDW fields.

Terms linear in derivative do occur by coupling to the E$_u$ bilinears. This coupling will shift the zero momentum $\Delta_0$ in the presence of LC. But as long as the dispersion is well approximated with a parabola, this shift will not change the dispersion around the stable point, and the response will remain isotropic. Therefore we will neglect these terms even in the presence of LC order.

\subsection{Expanded action}

We find the expanded action (\ref{eq_action_sc}) in terms of the irreducible representation discussed above as $S=S_{\text{PDW}}+S_{0} + S_{\text{PDW}-0}$ where $S_0$ contains the homogeneous superconducting field
\begin{equation}
    S_0 = \int \limits_{\vec{x}} \kappa |\vec{D} \Delta_{0}|^2+r_0 |\Delta_0|^2 + \frac{u}{2}|\Delta_0|^4 \, ,
\end{equation}
$S_{\text{PDW}}$ the PDW fields, and $S_{\text{PDW}-0}$ their interaction. $S_{\text{PDW}}$ can be divided further, $S_{\text{PDW}}=S_{\sectA}+S_{\sectB } + S_{\sectA-\sectB}$, one for each sector respectively, which will take the same form, but with independent parameters
\begin{widetext}
	\begin{equation}
	\begin{split}
%	\Big[ (D_x \Delta_{0})^{*}(D_x \Delta_{0}) + (D_y \Delta_{0})^{*}(D_y \Delta_{0}) \Big]
	%S_0 &= \int \limits_{\vec{x}} \kappa |\vec{D} \Delta_{0}|^2+r_0 |\Delta_0|^2 + \frac{u}{2}|\Delta_0|^4 \, ,\\
	%
	%
	S_\sectA&= \int \limits_{\vec{x}} \!\!
	\sum_{\ve{q}={\LCpm \ve{Q}_{1,2}}}  \!\!\!\kappa_1 |\vec{D} \Delta_{\ve{q}}|^2
	%\Big[ \sum_{q={\pm \QAx, \pm \QAy}}(D_x \Delta_{q})^{*}(D_x \Delta_{q}) + (D_y \Delta_{q})^{*}(D_y \Delta_{q}) \Big] 
	+ 
	 \kappa_2 
	 \Big[  \sum_{\ve{q}={\LCpm \QAx}}\!\!\! \Big( |D_x \Delta_{\ve{q}}|^2 \LCm |D_y \Delta_{\ve{q}}|^2 \Big)
	 - \!\!\!\!\!\!\sum_{\ve{q}={\LCpm \QAy}} \!\!\!\Big( |D_x \Delta_{\ve{q}}|^2 \LCm |D_y \Delta_{\ve{q}}|^2 \Big) \Big] 
%	 \Big[  \sum_{\ve{q}={\pm \QAx}} \Big( (D_x \Delta_{\ve{q}})^{*}(D_x \Delta_{\ve{q}}) - (D_y \Delta_{\ve{q}})^{*}(D_y \Delta_{\ve{q}}) \Big) - \sum_{\ve{q}={\pm \QAy}} \Big( (D_x \Delta_{q})^{*}(D_x \Delta_{\ve{q}}) - (D_y \Delta_{\ve{q}})^{*}(D_y \Delta_{\ve{q}}) \Big)  \Big] 
	%
	 +\!\!\! \sum_{\ve{q}={\LCpm \ve{Q}_{1,2}}}\!\!\!r|\Delta_{\ve{q}}|^2 
	%+ r(|\Delta_{\QAx}|^2 + |\Delta_{\QAy}|^2 + |\Delta_{-\QAx}|^{2} + |\Delta_{-\QAy}|^2 ) 
	+ \frac{u_0}{2}  |\Delta_{\ve{q}}|^4 \\
	&
	%+ \frac{u_0}{2}  (|\Delta_{\QAx}|^2+ |\Delta_{-\QAx}|^{2}  + |\Delta_{\QAy}|^2 + |\Delta_{-\QAy}|^2 )^2 \\
	+ \frac{u_1}{2} \Big[|\Delta_{\QAx}|^2+ |\Delta_{-\QAx}|^2 - |\Delta_{\QAy}|^2  - |\Delta_{-\QAy}|^2 \Big]^2
	+ \frac{u_2}{2}  \Big[ (|\Delta_{\QAx}|^2 - |\Delta_{-\QAx}|^2)^2 +(|\Delta_{\QAy}|^2 - |\Delta_{-\QAy}|^2)^2 \Big] \\
	&\\
	S_{\sectA-\sectB}&= \int \limits_{\vec{x}} v_{0}\Big[\sum_{\ve{q}={\LCpm \ve{Q}_{1,2}}}|\Delta_{\ve{q}}|^2  \Big]\Big[\sum_{\ve{q}'={\LCpm \overline{\ve{Q}}_{1,2}}}|\Delta_{\ve{q}'}|^2  \Big] 
	+\frac{v_{1}}{\sqrt{2}} \Big(|\Delta_{\QAx}|^2 \LCm |\Delta_{\LCm\QAx}|^2, |\Delta_{\QAy}|^2 \LCm |\Delta_{\LCm\QAy}|^2 \Big) \\
	&
\cdot\Big(|\Delta_{\QDx}|^2 \LCm |\Delta_{\LCm\QDx}|^2\LCm|\Delta_{\QDy}|^2 \LCp |\Delta_{\LCm\QDy}|^2, |\Delta_{\QDx}|^2 \LCm |\Delta_{\LCm\QDx}|^2\LCp|\Delta_{\QDy}|^2 \LCm |\Delta_{\LCm\QDy}|^2 \Big) \\
&\\
S_{\text{PDW}-0} &=  \int \limits_{\vec{x}} \gamma_{0} |\Delta_0|^2 \Big[\sum_{\ve{q}={\LCpm \ve{Q}_{1,2}}}|\Delta_{\ve{q}}|^2  \Big]+ \gamma_{1} ( |D_x \Delta_{0}|^2 - |D_y \Delta_{0}|^2)   (|\Delta_{\QAx}|^2+ |\Delta_{-\QAx}|^2 - |\Delta_{\QAy}|^2  - |\Delta_{-\QAy}|^2)\\ 
  \,    \\
	\end{split}
	\label{eq_full_action}
	\end{equation}
\end{widetext}	
%\com{$u_0=u(Q_x,Q_x,Q_x,Q_x)$$u(q_1,q_2,q_3,q_4) = \int_x u(x_1,x_2,x_3,x_4) e^{iq_1 x_1}e^{-iq_2 x_2}e^{iq_3 x_3}e^{-iq_4 x_4}$}
%
where we absorbed a factor $1/T$ in all coefficients. As stated previously, the coefficients could be traced back to the exact form of the full Ginzburg-Landau model, \ref{eq_action_sc}, but we will consider them as independent parameters \footnote{E.g. for a local interaction $U(\ve{q}_1,\ve{q}_2,\ve{q}_3,\ve{q}_4) = U\delta(\ve{q}_1-\ve{q}_2+\ve{q}_3-\ve{q}_4)$  we find $u_0 = 7U/4,u_1 =-U/4,u_2=-U/2$.}.

$S_{\sectB }$ has the same form as $S_{\sectA }$ with $\ve{Q}_{1,2} \rightarrow \ov{\ve{Q}}_{1,2}$ and $x,y \rightarrow \ov{x},\ov{y}$ where $\ov{x}=\frac{x+y}{\sqrt{2}},\ov{y}=\frac{-x+y}{\sqrt{2}}$. When in need of specifying both sectors we will append the subscript A or B to $r,u_0,u_1,u_2,\gamma_{0},\gamma_1$. The second-order term coefficient $r, r_0$ are assumed to be proportional to the temperature $T$, changing sign at the mean-field transition temperature $r \propto T-T_{\text{PDW}},r_0 \propto T-T_{\text{SC}}$. 

The interaction of the two sectors $S_{\sectA-\sectB}$ occurs in fourth-order terms and only involve the A$_{1g}$ and E$_u$ representation of both sectors. Throughout the text, we will only explicitly assume a stable A sector, meaning that \eqref{eq_dispersion} only supports local minima of momenta along the axis, for which $S_{\sectB}$ and $S_{\sectA-\sectB}$ drops out. However, we will reinsert the B sector when the result is directly generalizable. Discussion regarding the inclusion of both the A and B sectors are found in Appendix \ref{section_coupling_axial_diagonal}.
As mentioned above, we have left out terms consisting of bilinears that transform non-trivially under $U(1)$: $\Delta_0^2$, $(\Delta_{\QAx}\Delta_{-\QAx}\pm\Delta_{\QAy}\Delta_{-\QAy}), (\Delta_{\QDx}\Delta_{-\QDx}\pm\Delta_{\QDy}\Delta_{-\QDy})$. These secondary order parameters refers to so-called 4e superconducting order\cite{berg2009charge}, which we neglect in subsequent analysis.
%that transforms in $A_1,A_1$ and $B_1$. This contributes to $g_1$ term which I haven't taken into account in my note below.

\section{Effective anisotropic superconductor \label{section_an_sc}}
Now we will discuss the fate of the superconducting order in the presence of PDW vestigial order without any specific assumptions about the underlying instability or parameter regime. (Exploration of the ME-PDW vestigial phases is left for Section \ref{section_mean_field_equations}.) In the absence of long-range PDW order, $\av{\Delta_{\ve{Q}}}=0$, it is straightforward to integrate the PDW fields out, leaving an action only dependent on the vestigial order parameters and SC $\Delta_0$.

We begin by promoting the secondary order parameters to independent fields, which we do by decoupling the fourth-order terms in \eqref{eq_full_action} using the  Hubbard-Stratonovich transformation
\begin{equation}
\begin{split}
&e^{- \int \limits_{\vec{x}} \Phi^{*}  \frac{M}{2} \Phi } = \int \scr{D} \Psi \, e^{\;\int \limits_{\vec{x}} \Psi^{*} \frac{M^{-1}}{2} \Psi -\Phi^{*} \cdot \Psi}  \, .
\end{split}
\label{eq_HS}
\end{equation}
Here $\Phi$ is a vector of the bilinears listed in Table \ref{table} and $\Psi$ the corresponding vector of the auxiliary field to decouple that bilinear. The matrix $M$ is inferred from \eqref{eq_full_action} and contains the coupling constants. We will assume only a stable A sector, yielding a diagonal $M$. We denote the auxiliary fields with $\psi, N_{x^2\! \LCm \!y^2},\vec{l}$, dropping the A subindex, corresponding to the bilinear they decouple (see Table \ref{table}). Using this transformation, we express the partition function as
\begin{equation}
Z = \int \scr{D}\{ \Delta_{\ve{Q}}\}\scr{D} \Delta_0 \scr{D} \{ \psi, N_{x^2\! \LCm \!y^2} , \vec{l} \}  \, e^{-S_{\text{eff}}} \, ,
\end{equation}
with the effective action given by
\begin{widetext}
\begin{equation}
\begin{split}
&S_{\text{eff}}(\{ \Delta_{\ve{Q}} \},\Delta_0,\psi,N_{x^2\! \LCm \!y^2},\vec{l})=\int \limits_{\vec{k}} \chi^{-1}_{0}(k) |\Delta_0|^2+\frac{u}{2}|\Delta_0|^4 +\chi_{x}^{-1}(\vec{k})\left(|\Delta_{\QAx}|^2+|\Delta_{-\QAx}|^2 \right)  
+\chi_{y}^{-1}(\vec{k})\left(|\Delta_{\QAy}|^2+|\Delta_{-\QAy}|^2\right)  \\
&
 +\int \limits_{\vec{x}} (\psi+\gamma_0 |\Delta_0|^2) \left(|\Delta_{\QAx}|^2+|\Delta_{-\QAx}|^2 +|\Delta_{\QAy}|^2+|\Delta_{-\QAy}|^2\right) 
 +  l_x \left(|\Delta_{\QAx}|^2-|\Delta_{-\QAx}|^2\right) +l_y\left(|\Delta_{\QAy}|^2-|\Delta_{-\QAy}|^2\right) \\
 &
  + \int \limits_{\vec{x}} \left( N_{x^2\! \LCm \!y^2} + \gamma_1(|D_x\Delta_{0}|^2-|D_y\Delta_{0}|^2) \right) \left(|\Delta_{\QAx}|^2+|\Delta_{-\QAx}|^2 -|\Delta_{\QAy}|^2-|\Delta_{-\QAy}|^2\right) 
  -%\int \limits_{\vec{x}} 
  \left( \frac{\psi^2}{2u_0} + \frac{N_{x^2\! \LCm \!y^2}^2}{2u_1} +  \frac{\vec{l}^2}{2u_2}  \right)\\
\end{split}
\label{eq_effective_action}
\end{equation}
\end{widetext}
where $\chi_{0}^{-1}=r_0+\kappa(k_x^2+k_y^2)$ and $\chi_{x,y}^{-1}=r+\kappa_1(k_x^2+k_y^2)  \pm \kappa_2(k_x^2-k_y^2)$. We have left out the gauge field $\vec{A}$ (absorbing it in the phase gradient), considering an extreme type-II superconductor, for which the electromagnetic field energy can be ignored. We will treat $\psi, N_{x^2\! \LCm \!y^2}, \vec{l}$ on a mean-field level and only keep the uniform component (i.e. $\psi(\vec{q})=\psi (2\pi)^d\delta(\vec{q})$ etc.). The composite field $\psi$ will always have a non-zero and positive expectation value since it describes the fluctuations of the PDW state,
\begin{equation}
\begin{split}
&\frac{\psi}{u_0} = \int \limits_{\vec{x}}\av{|\Delta_{\QAx}|^2+|\Delta_{-\QAx}|^2 +|\Delta_{\QAy}|^2+|\Delta_{-\QAy}|^2} , \\
\end{split}
\label{eq_expectation_bilinear_0}
\end{equation} 
and is, therefore, not an order parameter. Similarly, developing a non-zero expectation value on any of the vestigial-orders parameters imply
\begin{equation}
\begin{split}
&\frac{N_{x^2\! \LCm \!y^2}}{u_1} =\int\limits_{\vec{x}}\av{|\Delta_{\QAx}|^2+|\Delta_{-\QAx}|^2 -|\Delta_{\QAy}|^2-|\Delta_{-\QAy}|^2}\\
&\frac{l_x}{u_2} =\int\limits_{\vec{x}}\av{|\Delta_{\QAx}|^2-|\Delta_{-\QAx}|^2}\\
&\frac{l_y}{u_2}  =\int\limits_{\vec{x}}\av{|\Delta_{\QAy}|^2-|\Delta_{-\QAy}|^2}\\
\end{split}
\label{eq_expectation_bilinear}
\end{equation} 
respectively. Even in absence of PDW order we find non-equivalent uniform static susceptibilities once the vestigial order parameters $N_{x^2\! \LCm \!y^2}, \vec{l}$ are finite
\begin{equation}
\begin{split}
&\chi_{\QAx}(0) = \frac{1}{r'+N_{x^2\! \LCm \!y^2}+l_x} , \quad \chi_{-\QAx}(0)  = \frac{1}{r'+N_{x^2\! \LCm \!y^2}-l_x} \\
& \chi_{\QAy}(0)  = \frac{1}{r'-N_{x^2\! \LCm \!y^2}+l_y} , \quad \chi_{-\QAy}(0)  = \frac{1}{r'-N_{x^2\! \LCm \!y^2}-l_y}  \\
\end{split}
\label{eq_static_succeptibilities}
\end{equation} 
here $r'=r+\gamma_0 |\Delta_0|^2 + \psi$ (see \eqref{eq_static_succeptibilities} for the full static susceptibilities). Without vestigial ordering, the transition temperature would be given by $r'=0$. Thus we see a splitting of the transition into the ordered PDW state and that the preemptive transition enhances the transition temperature. 

\subsection{Superconducting action}
The effective action \eqref{eq_effective_action} can be written on the form
\begin{equation}
\begin{split}
& S_{\text{eff}}(\{ \Delta_{\ve{Q}} \},\Delta_0,\psi,N_{x^2\! \LCm \!y^2},\vec{l})=\\
&
-V\left( \frac{\psi^2}{2u_0} +\frac{N^2_{x^2\! \LCm \!y^2}}{2u_1} + \frac{\vec{l}^2}{2u_2}  \right)
+\int_k \chi^{-1}_0(k) |\Delta_0|^2  + \int_r\frac{u}{2} |\Delta_0|^4   \\
&
+\int \limits_{k}  [\Delta_{\QAx} \Delta_{-\QAx} \Delta_{\QAy} \Delta_{-\QAy}]^{*}_i \G^{-1}_{i}  [\Delta_{\QAx} \Delta_{-\QAx} \Delta_{\QAy} \Delta_{-\QAy}]_i
\end{split}
\label{eq_effective_action_appendix_1}
\end{equation}
%
%\com{Here $N_{x^2\! \LCm \!y^2}=\frac{1}{L^d} \int_x N_{x^2\! \LCm \!y^2}(x)$ in the last integral}
where $V$ is the volume, and we have moved to momentum representation.

The kernel $\scr{G}$ is given by
\begin{equation}
\begin{split}
\G^{-1}_{1}(k) &= \chi_{x}^{-1}(k)+\psi+ N_{x^2\! \LCm \!y^2} + l_{x} + \Gamma_{x} \\
\G^{-1}_{2}(k) &= \chi_{x}^{-1}(k)+\psi + N_{x^2\! \LCm \!y^2}- l_{x}+  \Gamma_{x} \\
\G^{-1}_{3}(k) &= \chi_{y}^{-1}(k)+\psi  - N_{x^2\! \LCm \!y^2} + l_{y} + \Gamma_{y} \\
\G^{-1}_{4}(k) &= \chi_{y}^{-1}(k)+\psi - N_{x^2\! \LCm \!y^2} - l_{y} + \Gamma_{y} \, .\\
\end{split}
\end{equation}
%
%where $\chi_{x,y}^{-1}=r+\kappa_1(k_x^2+k_y^2)  \pm \kappa_2(k_x^2-k_y^2)$ 
%
Integrating over the PDW fields in \eqref{eq_effective_action} 
%we integrate them out assuming $\av{\Delta_{\ve{Q}}}=0$. 
we arrive at the effective action for the vestigial and homogeneous superconducting fields alone
\begin{equation}
\begin{split}
e^{-S_{\text{eff}}(\Delta_0,\psi,N_{x^2\! \LCm \!y^2},\vec{l})}&= \int \scr{D} \{ \Delta_{\ve{Q}} \} e^{-S_{\text{eff}}(\{ \Delta_{\ve{Q}} \},\Delta_0,\psi,N_{x^2\! \LCm \!y^2},\vec{l}) } \,. \\
\end{split}
\end{equation}
The new action takes the form 
\begin{equation}
\begin{split}
& S_{\text{eff}}(\Delta_0,\psi,N_{x^2\! \LCm \!y^2},\vec{l})=\\
&
-V\left( \frac{\psi^2}{2u_0} +\frac{N_{x^2\! \LCm \!y^2}^2}{2u_1} + \frac{\vec{l}^2}{2u_2}  \right)
+\int \limits_{\vec{k}} \chi^{-1}_0(\vec{k}) |\Delta_0|^2  + \int \limits_{\vec{x}} \frac{u}{2} |\Delta_0|^4   \\
&
+V\int \limits_{\vec{k}} \ln \Big[ \left( (\chi_{x}^{-1}(\vec{k})+\psi+ N_{x^2\! \LCm \!y^2}+ \Gamma_{x})^2 - l_{x}^2  \right) \\
&\times \left( (\chi_y^{-1}(\vec{k})+\psi- N_{x^2\! \LCm \!y^2}+ \Gamma_{y})^2 - l_{y}^2  \right) \Big]  \, ,\\
\end{split}
\label{eq_S_mean_field}
\end{equation}
where 
\begin{equation}
\begin{split}
\Gamma_{x,y} &= \frac{\gamma_0}{V}\int \limits_{\vec{k}} |\Delta_0(\vec{k})|^2 \pm \frac{\gamma_1}{V}\int \limits_{\vec{k}}(k_x^2-k_y^2)|\Delta_0(\vec{k})|^2
\end{split}
\end{equation}
%
%$\Gamma_{x,y}=\Gamma_{x,y}(\Delta_0)$ are functionals of the superconducting field, presented in \eqref{eq_Gamma_functionals_appendix} in Appendix \ref{appendix_effective_action}.
%
are functionals of the superconducting field. We expand the action in $\psi, N_{x^2\! \LCm \!y^2}, \vec{l}$ and $\Delta_0$ around their mean-field values (see \eqref{eq_mean_field_1})
\begin{equation}
\begin{split}
S_{\text{eff}}(\Delta_0,\psi,N_{x^2\! \LCm \!y^2},\vec{l}) \approx  S_{0}+ S_{\text{SC}}(\Delta_0)\, ,
 \end{split}
\end{equation}
$S_0 = S_{\text{eff}}|_{\text{MF}}$ is the action at the mean-field solution, and $S_{\text{SC}}$ the effective superconducting action. Above the superconducting transition temperature ($\av{\Delta_0}=0$) we find 
\begin{equation}
\begin{split}
& S_0 /V=
-\left( \frac{\psi^2}{2u_0} +\frac{N^2_{x^2\! \LCm \!y^2}}{2u_1} + \frac{\vec{l}^2}{2u_2}  \right)\\
&
+\int \limits_{\vec{k}} \ln \Big[ \left( (\chi_{x}^{-1}(\vec{k})+\psi+ N_{x^2\! \LCm \!y^2})^2 - l_{x}^2  \right) \\
&\times \left( (\chi_y^{-1}(\vec{k})+\psi- N_{x^2\! \LCm \!y^2})^2 - l_{y}^2  \right) \Big]  \, .\\
\end{split}
\label{eq_action_2D}
\end{equation}
%
%After expanding around $\Delta_0=0$ and 

In real space the effective superconducting action takes the form 
\begin{equation}
\begin{split}
& S_{\text{SC}}(\Delta_0) = \\
& \int \limits_{\vec{x}} r'_0|\Delta_0(\vec{x})|^2 + \left( \frac{1}{2m_{\text{p}}} \right)_{ij} (D_i\Delta_0(\vec{x}))(D_j\Delta_0(\vec{x}))^{*} \, ,\\
& \left( \frac{1}{2m_{\text{p}}} \right)_{ij} = \frac{\delta_{ij}}{2m_{\text{p},0}} + S_{ij} , \\
& S=
\begin{bmatrix}
\frac{\gamma_{1,\sectA} }{u_{1,\sectA}}N_{x^2-y^2} & \frac{\gamma_{1,\sectB} }{u_{1,\sectB}}N_{xy}\\
\frac{\gamma_{1,\sectB} }{u_{1,\sectB}}N_{xy}& -\frac{\gamma_{1,\sectA} }{u_{1,\sectA}}N_{x^2-y^2} \\
\end{bmatrix} \, .\\
\end{split}
\label{eq_effective_superconductive_action}
\end{equation}
Here we have reintroduced both the A and B sectors and used the mean-field equations \eqref{eq_mean_field_1} to identify the order-parameters. (The mean-field equations for the two primary nematic orders remain unaltered even in the presence of both A and B sector, see Appendix \ref{section_coupling_axial_diagonal}.)

The superconducting effective action \eqref{eq_effective_superconductive_action} is expressed in terms of an anisotropic pair-mass (with $\kappa = (2m_{\text{p},0})^{-1}$), induced by the nematic order parameters through the trace-less symmetric matrix $S$. (Here we have only explicitly included the primary nematic fields, that are linear in the PDW fluctuations $|\Delta_\ve{Q}|^2$.) This anisotropic pair mass is equivalent to an anisotropic stiffness, that may be observed for example as an anisotropy of the in-plane penetration depth in the supercondcuting state, or through an anisotropy of the near $T_c$ normal state conductivity due to superconducting fluctuations \cite{wardh2019inprep}.

The coupling $r'_0=  r_0+\frac{\gamma_{0,\sectA}}{u_{0,A}}\psi_{\sectA}+\frac{\gamma_{0,\sectA}}{u_{0,\sectB}}\psi_B$ is the renormalized inverse static susceptibility. Since $\Delta_0$ is single component, and its amplitude is rotationally symmetric, it cannot couple directly to the nematic order, as seen from the fact that only the symmetric PDW fluctuations $\psi_{\sectA/\sectB}$ contribute. As discussed in Section \ref{section_derivative}, the expression \eqref{eq_effective_superconductive_action} is expected to hold even in the presence of LC order, although the superconductor would acquire a small finite momentum. 

\subsection{Enhancement of anisotropic superconducting fluctuations  near PDW instability\label{section_super_fluct}\label{section_eff_SC}}

In \eqref{eq_effective_superconductive_action} we have found an effective superconducting action, renormalized by the PDW fluctuations and the possible vestigial nematic order parameters $N_{xy}$ and $N_{x^2-y^2}$. Note that the superconductor is anisotropic in its dynamics, and at this level, the static order parameter is still isotropic. However, in the ordered SC state, the nematic order will affect the order parameter. In assuming d-wave superconductivity, there is a coupling $\Delta_d\Delta_s^*N_{x^2-y^2}+h.c.$ (not considered here), which will induce a subleading s-wave component, effectively shifting the gap nodes. Nevertheless, as we argue below, for a superconductor with low phase stiffness, the effect of even a weak nematic field on the fluctuations may be dramatic, even though the effect on the static gap may be small.

As a digression, we note that this scenario of the effect of vestigial nematic order from superconducting fluctuations on the superconductor is similar in spirit but also different from electron-doped Bi$_2$Se$_3$. The latter has a multicomponent SC order parameter, which may itself form a vestigial nematic phase, which in turn would also affect the dynamics in the normal state\cite{hecker2018vestigial}. In our case, instead, it is the finite momentum (PDW) superconducting components that give rise to the vestigial nematic order.     

One way to probe the anisotropic stiffness of the superconductor is to study the contribution to the conductivity from superconducting fluctuations above $T_c$, the paraconductivity. At $T_c$, this contribution will diverge, reflecting the lifetime of Cooper pairs, and we expect to see a strong signature of the nematic order near $T_c$. As derived in \cite{wardh2019inprep}, the Aslamazov-Larkin expression for the in-plane paraconductivity of a layered superconductor (inter-layer distance $d$) with anisotropic stiffness is given by 
\begin{equation}
\begin{split}
&\overline{\sigma }_{\text{p}} =\frac{e^2}{16 \hbar d}\frac{1}{(T/T_c)-1}\sqrt{\det\left(\overline{m}_{\text{p}}\right)}\overline{m}_{\text{p}}^{-1}\\
&\underrel{\text{princ.}}{=}
\frac{e^2}{16 \hbar d}\frac{1}{(T/T_c)-1}
\begin{bmatrix}
\sqrt{m_{\text{p},b}/m_{\text{p},a}} & 0 \\ 
0 &\sqrt{m_{\text{p},a}/m_{\text{p},b}} \\ 
\end{bmatrix}
\\
\end{split}
\end{equation}
where $a,b$ refers to the principal axes of the conductivity, such that the last expression holds in the principal frame. Given a nematic distortion, in the form of \eqref{eq_effective_superconductive_action}, the pair-mass quotient is given by 
\begin{equation}
\frac{m_{\text{p},b}}{m_{\text{p},a}} = \frac{1/2m_{\text{p},0} + \sqrt{S_{xx}^2+S_{xy}^2}}{1/2m_{\text{p},0} - \sqrt{S_{xx}^2+S_{xy}^2}} \, ,
\label{eq_mass_anisotropy}
\end{equation}
where the angle of the $a$ axis (corresponding to the axis of highest conductivity) to the crystal $x$-axis is given by 
%\begin{equation}
%    \theta = \arccos %(\frac{1}{2}+\frac{S_{xx}}{2\sqrt{S_{xx}^2+S_{xy}^2}} )^{1/2}\,.
%\end{equation}
\begin{equation}
    \theta = \arctan \frac{\sqrt{S_{xx}^2+S_{xy}^2}-S_{xx}}{S_{xy}}\,.
\end{equation}

Thus, in the presence of both $N_{xy}$ (i.e.\ $S_{xy}\neq 0$) and $N_{x^2\! \LCm \!y^2}$ (i.e.\ $S_{xx}\neq 0$) the principal axes of conductivity will not be aligned with the symmetry axes of the crystal, and will rotate if the relative amplitude of the two fields change with temperature.

%\subsection{Enhancement of anisotropic superconducting fluctuations  near PDW instability\label{section_super_fluct}}
There is, in fact, evidence for highly anisotropic superconducting fluctuations in transport measurements done on thin films of underdoped LSCO \cite{wu2017spontaneous,wardh2019inprep,bovzovic2023nematicity}, consistent with a high pair-mass ratio $m_{\text{p},b}/m_{\text{p},a}$. This ratio increases as the underdoped critical point is approached, while the quotient of normal masses $m_{b} / m_{a}$ remains near 1 \cite{wardh2019inprep}. The crystals show very weak signs of lattice distortion, remaining effectively tetragonal, which is in line with the development of electronic nematicity coupling directly to the superconductor, and not through strain\cite{wu2017spontaneous,bovzovic2023nematicity}, consistent with the anisotropic superconductor described in \eqref{eq_effective_superconductive_action}. In addition, the principal axes of the paraconductivity (seen close to $T_c$) and the normal conductivity are in general not aligned with each other, or with the crystal axes, which is consistent with the presence of both $B_{1g}$ and $B_{2b}$ nematic order. 

Nevertheless, the analysis leading up to \eqref{eq_effective_superconductive_action} does not by itself explain why the superconducting stiffness-anisotropy would be enhanced compared to other observables that couple to nematicity, such as the normal electron conductivity, orthorhombic lattice distortions, and the superconducting gap (as discussed above). 
However, a natural explanation for this is evident in the expression for the pair-mass ratio \eqref{eq_mass_anisotropy}; if the isotropic pair mass $m_{\text{p},0}$ is sufficiently large (i.e.\ stiffness small), the quotient $m_{\text{p},b}/m_{\text{p},a}$ will become large even for a  small nematic tensor $S$. Without an explicit microscopic model of how the nematic order couples to normal and paired electrons this is only a qualitative statement, but that the phase stiffness is small in the underdoped cuprates is well-established\cite{emery1995importance}. 

In fact, proximity to a PDW instability provides a unified conceptual framework in which both the low stiffness and the more recently observed nematic distortion thereof can be understood.  As discussed in Section \ref{section_model}, and more detailed in \cite{waardh2018suppression}, such an instability is expected to influence also the uniform component by deforming the spectrum of superconducting fluctuations giving a large effective pair mass. In other words, the availability of low energy finite momentum pair excitations suppresses the stiffness to real space deformations. Also, as we will further elucidate in the next section, the fluctuations of a (metastable) PDW state can generate vestigial nematic order that acts to deform the stiffness. Thus, approaching the finite momentum instability provides a mechanism for generating highly anisotropic superconducting fluctuations, both through creating a low phase-stiffness, yielding a high susceptibility towards an anisotropic distortion, as well as providing the distortion itself.
This scenario is depicted in Figure \ref{fig_pahase_diagram}, where the pseudogap is made up of vestigial phases set up by an underlying PDW state (possibly ME-PDW, as discussed in the following sections), with anisotropic superconducting fluctuations.

\section{Intertwined nematic and loop-current orders in the vestigial ME-PDW phase \label{section_enhanced_sym}\label{section_mean_field_equations}}
In this section we will investigate the various possible vestigial phases from an action of the form \eqref{eq_full_action}. Specifically we are interested in the nature of generation of the nematic order coupling to the dynamic response of the homogeneous superconductor through the action \eqref{eq_effective_superconductive_action}.

\begin{figure}
	\centering
	\includegraphics[width=0.48\textwidth]{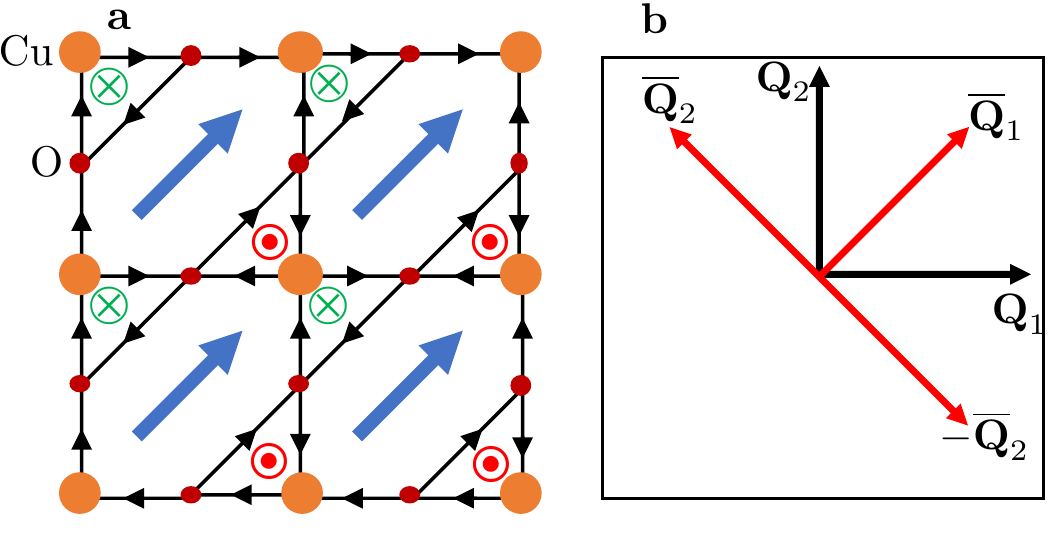} 
	\caption{\label{fig_loop_current_PDW_analogy} {\bf a}
		A loop-current state with microscopic circulating currents suggested to account for the observed time-reversal symmetry seen in cuprates (the so-called $\Theta_2$ state)\cite{simon2002detection,kaminski2002spontaneous}. The LC order is shown as a blue arrow. {\bf b} The corresponding $xy$ ME-PDW state. %Black (red) arrows corresponds to momenta in the A (B) sector. 
	    Black arrows correspond to stable {\em axial} PDW components, referred to as the A sector, while red arrows show an alternative configuration for stable {\em diagonal} PDW components (B sector). }
\end{figure}

The ME-PDW state argued to be consistent with polarized ARPES measurements\cite{kaminski2002spontaneous} has $xy$ LC order corresponding to $(\Delta_{\QAx},\Delta_{-\QAx},\Delta_{\QAy},\Delta_{-\QAy})=(\Delta,0,\Delta,0)$, which we will refer to as $xy$ ME-PDW. In Figure \ref{fig_loop_current_PDW_analogy}a the PDW momenta for the $xy$ ME-PDW state is shown alongside its circulating current analogue\cite{simon2002detection}.
%has the symmetry direction aligned with the $xy$ diagonal, yielding an expectation value of the subleading B$_{2g}$ nematic order parameter $l_{x}l_y \neq 0$. 
As a speculative scenario for the cuprates, we will consider ME-PDW as a mean-field ground state for the pseudogap phase. 
%
%So far, we have not considered the stability of the various vestigial phases of \eqref{eq_full_action}. 
%It is instructive to first consider the mean-field ground states for $r<0$ of $S_A$ in \eqref{eq_full_action}. 
The overall stability of the action \eqref{eq_full_action} requires $u_0>0$. When $u_1>0,u_2>0$ no point-group symmetry is broken. %For $u_1<0,u_2>0$ we have nematic order without LC order (this case is discussed in Appendix \ref{appendix_negative_u1}), while $u_2<0$ is necessary for LC order. 
For $u_1<0$ we have the possibility for nematic order without LC order, % (for completeness, this case is discussed in Appendix \ref{appendix_negative_u1}),
while $u_2<0$ is necessary for LC order. 
%(Following results holds equally for stable sector B and unstable sector A, taking into account the $45^{\circ}$ shift.) 
% A finite LC order, which breaks the $D_{4\text{h}}$ symmetry to $C_{2v}$, 
%
%\subsection{The mean-field ME-PDW ground state}
%
The mean-field solutions of $S_A$ in \eqref{eq_full_action}, and the corresponding stable vestigial phase, for $u_{2}<0, u_1>0$
%, assuming un-coupled sectors $v_{0,1}=0, \gamma_{0,(\sectA,\sectB)}=0$ 
are presented in Table \ref{ground_state_V2}. Here and subsequently, we use the notation
\begin{equation}
\alpha = \frac{u_0}{|u_2|}, \beta= \frac{u_1}{|u_2|}\,,
\end{equation}
which parameterize the relative repulsion strength of fluctuation and primary nematic field. (The stability of the action \eqref{eq_full_action} requires $\alpha>0.5$ for $\beta>0.5$ and $\alpha>1-\beta$ for $\beta<0.5$.)
%
%The ME-PDW ground states with LC order has reduced $D_{4\text{h}}$ symmetry to $C_{2v}$. 
%

\begin{table*}
\begin{tabularx}{1.0\hsize}{lcccc}
 \hline \hline
\begin{tabular}{l} Parameter regime \end{tabular}& \begin{tabular}{l} PDW Ground state \\ /Nematic order \\ (sub.) \end{tabular} &\begin{tabular}{l}PDW 1st excited  \\ /Nematic order \\ (sub.)  \end{tabular} & \begin{tabular}{l}Stable vestigial phase \end{tabular} & \begin{tabular}{l} Transition order \\ from normal state \end{tabular}\\ \hline
\begin{tabular}{l} $0 <  \beta < 0.25$  ,\\   $1-\beta < \alpha$ \end{tabular}	&\begin{tabular}{c}$(\Delta,0,0,0)$ \\ $x^2-y^2$ \end{tabular}&   \begin{tabular}{c}$(\Delta,0,\Delta,0)$ \\ $xy $ \end{tabular}  & $x^2\!-\!y^2$ LC & 1st\\ \hline
\begin{tabular}{l}$0.25< \beta < 0.5 $   ,\\    $1\!-\!\beta < \alpha < \!\frac{\beta}{4 \beta -1}$ \end{tabular} &\begin{tabular}{c} $(\Delta,0,0,0) $\\ $x^2-y^2$ \end{tabular}&  \begin{tabular}{c} $(\Delta,0,\Delta,0)$ \\$xy$ \end{tabular} & $x^2\!-\!y^2$ LC & 1st \\ \hline
%
%\begin{tabular}{cc} $1-\beta < \alpha < \frac{2+\beta}{4\beta-1}$ & 1st \\ $ \frac{2+\beta}{4\beta-1}<\alpha $ & 2nd \end{tabular}\\ \hline
%
\begin{tabular}{l} $0.25< \beta < 0.5$   ,\\   $\frac{\beta}{4\beta -1} <\alpha $\end{tabular} &\begin{tabular}{c}$(\Delta,0 ,\Delta',\Delta')$ \\ $x^2-y^2 $\end{tabular}&\begin{tabular}{c}   $(\Delta,0,0,0)$  \\  $x^2-y^2 $ \end{tabular}& $x^2\!-\!y^2 $ LC & \begin{tabular}{cc} $\frac{\beta}{4\beta-1} < \alpha < \frac{2+\beta}{4\beta-1}$ & 1st \\ $ \frac{2+\beta}{4\beta-1}<\alpha $ & 2nd \end{tabular}	\\	\hline
\begin{tabular}{l} $0.5< \beta < 1 $  ,\\   $0.5 < \alpha < \beta $\end{tabular} &\begin{tabular}{c}$(\Delta,0,\Delta,0)$ \\ $xy$ \end{tabular}&\begin{tabular}{c}  $ (\Delta,0 ,\Delta',\Delta') $  \\  $x^2-y^2 $ \end{tabular} & $xy$ LC& 1st\\ 	\hline
\begin{tabular}{l} $0.5< \beta < 1$   ,\\   $\beta < \alpha$ \end{tabular} &\begin{tabular}{c}$(\Delta,0,\Delta,0)$ \\ $xy$ \end{tabular}&\begin{tabular}{c}   $(\Delta,0 ,\Delta',\Delta') $ \\  $x^2-y^2$  \end{tabular} & \Big( \begin{tabular}{c}  
	Low temp. $xy$ LC  \\ 
	 High temp. $x^2-y^2$ LC\end{tabular}\Big) & \begin{tabular}{cc} $\beta < \alpha < \frac{2+\beta}{4\beta-1}$ & 1st \\ $ \frac{2+\beta}{4\beta-1}<\alpha $ & 2nd \end{tabular} \\	\hline
\begin{tabular}{l} $1 < \beta   $,\\  $ 0.5 < \alpha$ \end{tabular} &\begin{tabular}{c}$(\Delta,0,\Delta,0)$ \\ $xy$ \end{tabular}&\begin{tabular}{c}  $ (\Delta,0 ,\Delta',\Delta')  $ \\ $ x^2-y^2 $ \end{tabular} & $xy$ LC & \begin{tabular}{cc} $0.5<\alpha<1$ & 1st \\ $1<\alpha$ & 2nd \end{tabular}\\	\hline \hline
\end{tabularx}
	\caption{\label{ground_state_V2} \label{table_preemtive_sol}ME-PDW mean-field ground and first excited states for $S_{\sectA }$ in \eqref{eq_full_action} alongside the possible vestigal phases and the corresponding transition order. Here $r<0$, $\alpha = \frac{u_0}{|u_2|}$, and $ \beta= \frac{u_1}{|u_2|}$.
	States are expressed in the form $(\Delta_{\QAx},\Delta_{-\QAx},\Delta_{\QAy},\Delta_{-\QAy})$ together with their subleading nematic order. $(\Delta,0,\Delta,0)$ corresponds to the $xy$ ME-PDW state.
	%This assumes $v_{0,1}=0, \gamma_{(\sectA,\sectB),0}=0$.
	(Equivalent table holds for sector \sectB{} for $(\Delta_{\QDx},\Delta_{-\QDx},\Delta_{\QDy},\Delta_{-\QDy})$ and $xy \leftrightarrow x^2-y^2$.)
	The stability of the vestigial phases assumes a sufficiently low temperature ($R$), see Figure \ref{figure_multi_beta_075}. The transition order refers to the vestigial to normal phase transition. For $0.5<\beta<1, \beta<\alpha$ the transition order between the low and high temperature phase is first order.
	}
\end{table*}
%
%\subsection{\com{Remove?}Disordering the ME-PDW mean-field solution --- the vestigial phase}
%

The mean-field state breaks both continuous $U(1)$ gauge symmetry and the discrete point-group symmetry simultaneously. In the vestigial phase, the point-group symmetry breaking preempts the continuous symmetry breaking. Given that fluctuations act to restore the continuous symmetry, we expect that the vestigial phase breaks the same point-group symmetry as the mean-field solution. %\cite{fernandes2019intertwined}.
From this line of reasoning we expect to find a vestigial $xy$ LC phase ($\vec{l}=(l,l)$ without long-range PDW order)
%$, N_{xy}\neq 0$, $N_{x^2-y^2} =0$) 
above the transition to the $xy$ ME-PDW. Surprisingly, for $0.5<\beta<1, \beta<\alpha$ we find that the $xy$ LC phase can become unstable to a $x^2\! - \!y^2$ LC phase ($\vec{l}=(l,0)$)
%$,N_{xy}= 0, N_{x^2-y^2} \neq 0$). 
at higher temperature (see Figure \ref{figure_multi_beta_075}). Thus, the mean-field ground state is preempted by a low-temperature vestigial phase, sharing the same $xy$ symmetry, and a high-temperature vestigial phase, with a different symmetry ($x^2-y^2$). %The reason for forgoing this result is that 
This possibility can be understood as a result of a fluctuation induced transition between the mean-field ground and first excited state, which are both listed in Table \ref{ground_state_V2}.
%
%The possibility will arise for $0.5<\beta<1, \beta<\alpha$, where $xy$ ME-PDW and $x^2-y^2$ ME-PDW are the ground and first excited state respectively.
%
%In \ref{sec_vestigial_mean_field} we discuss the details of how the Table \ref{ground_state_V2} was obtained.
%
Near this transition we find a state with {\it soft} nematic order, which is discussed in Section \ref{subsection_enhanced_sym}
and \ref{section_soft}) 

In the continuation of this section we derive the content of Table \ref{ground_state_V2} and study the phase diagram for $0.5<\beta<1, \beta<\alpha$, presented in Figure \ref{figure_multi_beta_075}. Some details are left for the Appendix \ref{sec_xy_LC}, \ref{sec_xx+yy_LC} and \ref{appendix_energy}.

\subsection{Note on primary and subleading nematic orders \label{sec_note}}

Again, for simplicity, we assume that only sector A is stable for the following development. However, it is important to note that the A and B sector supports different {\it primary} nematic fields, $N_{x^2\! - \! y^2}$ and $N_{xy}$ respectively, while both supports the {\it subleading} nematic order $l_x^2-l_y^2$ and $l_x l_y$. A finite LC order implies subleading nematic order $l_x^2-l_y^2$, $l_x l_y$ transforming as B$_{1g}$ and B$_{2g}$, respectively. The subleading nematic orders are to fourth order in PDW fields, while the primary nematic fields $N_{x^2\! - \! y^2},N_{xy}$ (listed in Table \ref{table}) are to second order in the PDW fields. Specifically, sector A only supports the primary nematic order $N_{x^2\! - \! y^2}$, but not $N_{xy}$. Thus, an $xy$ LC order, $\vec{l}=(l,l)$, implies subleading B$_{2g}$ nematic order, $l_xl_y$, but no primary, $N_{xy}$. In contrast, an $x^2 \! - \!y^2$ LC order implies both secondary and primary B$_{1g}$ order, $l_x^2-l_y^2, N_{x^2\! - \! y^2}$. (The reverse is true for the B sector.)
%
% Moved down again
%Only considering the A sector we do not find any primary ($N_{xy}$) B$_{2g}$ nematic order for the $l_x=l_y$ solution. However, if both the A and B sector were stable, the subleading ($l_xl_y$) B$_{2g}$ nematic order would induce a finite primary B$_{2g}$ nematic order $N_{xy}$ (supported by sector B) through the coupling set by $v_{1}$ in \eqref{eq_full_action}. In Appendix \ref{section_coupling_axial_diagonal} we discuss the inclusion of both sectors, but for the present discussion this will be implicit.

%
%\subsection{Vestigial mean-field equations}
%
\subsection{Vestigial mean-field solutions \label{sec_vestigial_mean_field}}

We will explore the possibility of vestigial-ordering by considering the mean-field solutions for $\psi, N_{x^2\! \LCm \!y^2},\vec{l}$, of the effective action \eqref{eq_action_2D}, given by the solutions to the mean-field equations $\frac{\p S_{\text{eff}}}{\p \Phi}=0 \Rightarrow \frac{\delta S_{\text{eff}}}{\delta \psi }=0,\frac{\delta S_{\text{eff}}}{\delta N_{x^2\! \LCm \!y^2}}=0$, $\frac{\delta S_{\text{eff}}}{\delta l_{x,y} }=0$
%. In the high-temperature limit, these are given by
%
\begin{equation}
\begin{split}
\psi &=u_0\int \limits_{\vec{k}} \chi_{\QAx}(\vec{k})+ \chi_{\LCm\QAx}(\vec{k})+ \chi_{\QAy}(\vec{k})+ \chi_{\LCm\QAy}(\vec{k})\\
N_{x^2\! \LCm \!y^2}&=u_1\int \limits_{\vec{k}} \chi_{\QAx}(\vec{k})+ \chi_{\LCm\QAx}(\vec{k})- \chi_{\QAy}(\vec{k})- \chi_{\LCm\QAy}(\vec{k})\\
l_x&=u_2\int \limits_{\vec{k}} \chi_{\QAx}(\vec{k})- \chi_{\LCm\QAx}(\vec{k})\\
l_y&=u_2\int \limits_{\vec{k}} \chi_{\QAy}(\vec{k})- \chi_{\LCm\QAy}(\vec{k})\\
\end{split}
\label{eq_mean_field_1}
\end{equation}
where
\begin{equation}
\begin{split}
&\chi_{\pm\QAx}(\vec{k}) = \frac{1}{\chi'_{x}(\vec{k})^{-1}+N_{x^2\! \LCm \!y^2}\pm l_x} \, ,\\
& \chi_{\pm\QAy}(\vec{k})  = \frac{1}{\chi'_{y}(\vec{k})^{-1}-N_{x^2\! \LCm \!y^2}\pm l_y} \\
 \end{split}
%\label{eq_dynamic_succeptibilities}
\label{eq_static_succeptibilities}
\end{equation} 
are the static susceptibilities with $\chi'_{x,y}(\vec{k})^{-1}=r'+\kappa_1(k_x^2+k_y^2)  \pm \kappa_2(k_x^2-k_y^2)$, $r'=r+ \psi + \gamma_0 |\Delta_0|^2$.

The most extreme example of a preemptive transition into a vestigial phase happens in 2D, where the integral for $\psi$ is infrared-divergent. This divergence leads to finite PDW susceptibilities \eqref{eq_static_succeptibilities}, implying no long-range PDW order. This is just a restatement of the Mermin-Wagner theorem: Continuous symmetries will not form long-range order at finite temperature in 2D. The vestigial order parameters, however, being discrete Ising-like orders can break the point-group symmetries. Continuing with the 2D case, we find the mean-field equations of \eqref{eq_action_2D} as
%(see also Appendix \ref{appendix_2D_energy_mean_field})
%
\begin{subequations}
	\label{mean}
	\begin{align}
	\label{mean:r}
	r' &= r^R \\
	\notag 
	&-\frac{u_0}{4\pi \ov{\kappa}} \ln[(r'+N_{x^2\! \LCm \!y^2})^2-l_x^2][(r'-N_{x^2\! \LCm \!y^2})^2-l_y^2] \\
	\label{mean:phim}
	N_{x^2\! \LCm \!y^2}&= - \frac{u_1}{4 \pi \ov{\kappa}} 
\ln \frac{(r'+N_{x^2\! \LCm \!y^2})^2-l_x^2}{(r'-N_{x^2\! \LCm \!y^2})^2-l_y^2}
	 \\
	 \label{mean:etax}
	l_{x}&= 
-\frac{u_2 }{4 \pi \ov{\kappa}}  
\ln \frac{r'+ N_{x^2\! \LCm \!y^2}+l_{x}}{r' + N_{x^2\! \LCm \!y^2}-l_{x}}  
	 \\
	 	 \label{mean:etay}
	l_{y}&= 
-\frac{ u_2 }{4 \pi \ov{\kappa}}  
\ln \frac{r'- N_{x^2\! \LCm \!y^2}+l_{y}}{r' - N_{x^2\! \LCm \!y^2}-l_{y}}  \, ,
	 \\
	 \notag
	\end{align}
\end{subequations}
%
%assuming $\Delta_0=0$. 
where we introduced $r^R=r + \frac{u_{0}}{\pi\ov{\kappa}}  \ln(\ov{\kappa}\Lambda^2)$, $\ov{\kappa}=\sqrt{|\kappa_1^2-\kappa_2^2|}$, and $\Lambda$ as the momentum cut-off. Instead of using $\psi$ we have expressed the mean-field equations in terms of 
%$r'=r+\gamma_0 \Delta_0^2+\psi$. 
$r'=r+\psi$, where we assume $\Delta_0=0$. This is natural since $\psi$ describes Gaussian fluctuations of the PDW fields, which renormalizes the bare static susceptibility $r^{-1}$. 

Care must be taken when considering solutions to \eqref{mean}. First, in the absence of long-range PDW order, only solutions fulfilling $r' + N_{x^2\! \LCm \!y^2} \pm l_{x}, r' - N_{x^2\! \LCm \!y^2} \pm l_{y}>0$ can be considered physical. Secondly, solutions to \eqref{mean} are generally singular, meaning that solutions with finite order do not generally coincide with solutions without order in the limiting cases. Therefore we will have to consider all possible combinations of ordering independently. In addition to the trivial normal state without any ordering we find the following solutions
\begin{enumerate}
\item {\bf $xy$ LC state:} LC ordered state with  $xy$ nematic order, $l_x=l_y \neq 0$. \footnote{Only considering the A sector we do not find any primary ($N_{xy}$) B$_{2g}$ nematic order, as discussed in \ref{sec_note}.}. Solutions are presented in Appendix \ref{sec_xy_LC}.
\item {\bf LC saddle-point solution:} Unstable LC ordered state with both $x^2\! \LCm \!y^2$ and $xy$ nematic order, $l_x \neq l_y \neq 0$. Solutions are presented in Appendix \ref{sec_xy_LC}.
\item {\bf $x^2\! \LCm \!y^2$ LC state:} LC ordered state with with $x^2\! \LCm \!y^2$ nematic order, $l_{x,y}\neq 0, l_{y,x} =0$ and $N_{x^2\! \LCm \!y^2}\neq0$. Solutions are presented in Appendix \ref{sec_xx+yy_LC}.
\end{enumerate}
Solutions with only nematic order and no LC order is of secondary interest and presented in Appendix \ref{appendix_nematic}, for completeness.

In finding the vestigial mean-field solutions it is convenient to re-scale the order parameters to unit less quantities, $ \tilde{l}_x = 2 \pi \ov{\kappa}l_x/|u_2|$, and equivalently for other variables. (See Appendix \ref{sec_xy_LC} and \ref{sec_xx+yy_LC} for details.) However, for notational clearness we will suppress the tilde even when the parameters should be interpreted as unit less. The susceptibility gains an additional shift
\begin{equation}
R  = \tilde{r}^R - 2\alpha \ln \left( \frac{|u_2|}{2 \pi \ov{\kappa}}  \right)  \, .
\end{equation}
 Here $R$ is assumed to be tunable with temperature through its dependence on the bare susceptibility $r$.

The mean-field solutions only guarantee local stability, and we must compare the absolute energy of the different phases in order to find the ground state. The energy is presented in Appendix \ref{appendix_energy}. The energy expression \eqref{energy_given_r} was used to find the stable vestigial phases listed in Table\ref{table_preemtive_sol}.
% MOVED
For $0<\beta<0.5$ ($1<\beta$), the $x^2-y^2$($xy$) LC state is the stable state, regardless of $\alpha$ (and for low enough $R$). In contrast, for $0.5<\beta<1$, there is a transition between the $x^2\! - \!y^2$ and $xy$ LC states for $\beta<\alpha$, as $R$ ($\propto T$) is lowered, while the $xy$ LC state is the only possible ordered state for $\alpha<\beta$. Thus, after including fluctuations, there is an induced transition between the would-be mean-field ground and first excited state (see Table \ref{table_preemtive_sol}), resulting in a high-temperature $x^2\! - \!y^2$ LC and low-temperature $xy$ LC phase, separated by a first-order transition.

\subsection{Phase diagram and the $x^2\! - \!y^2$ and $xy$ LC transition \label{subsection_enhanced_sym}\label{section_comp}}
\begin{figure*}
	\includegraphics[width=1.0\textwidth]{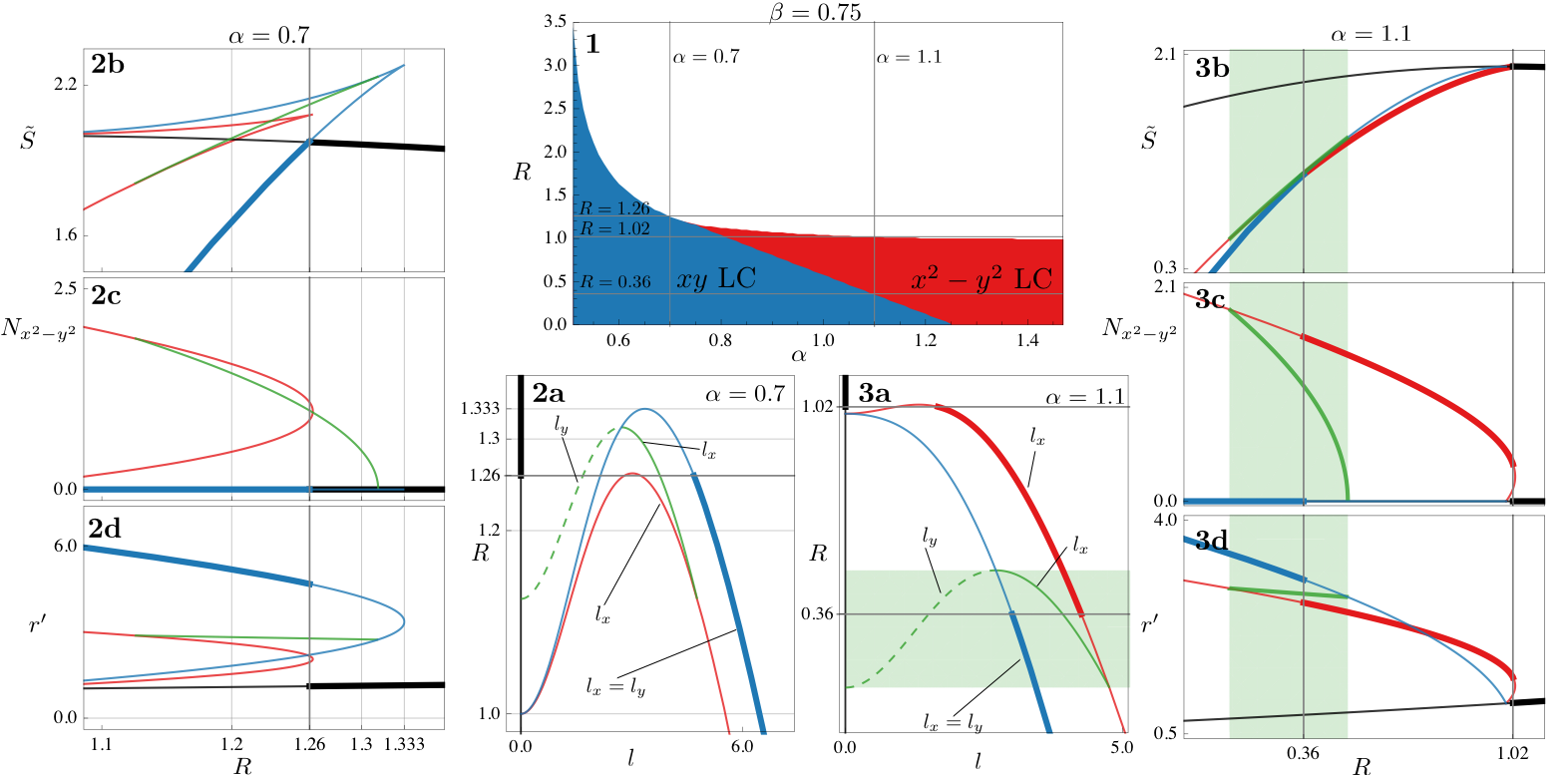} 
	\caption{\label{figure_multi_beta_075}  \footnotesize
		Two solved systems for $\beta=0.75$ with three metastable phases: Normal phase with no order (black lines), $xy$ LC phase $l_x=l_y$ (blue) and $x^2-y^2$ LC phase $l_x>0,l_y=0$ (red).
		{\bf 1)} Phase diagram as a function of $\alpha$ and $R$ ($R=0$ is an arbitrary starting point). 
		%For $0.4\lesssim\alpha<0.5$ there is support for a $l$-axis phase with nematic order as the temperature is lowered. For $0.5<\alpha<0.75$ the stable phase is a $xy$ LC phase without nematic order. For $0.75 < \alpha \lesssim 0.125$ there exists a transition between the $l$-axis and diagonal LC.
		%
		{\bf 2)} Developing of order for $\alpha = 0.7$. Thick lines corresponds to the global stable phase (in {\bf 1}) while thin lines corresponds to local extrema of the action. 
		Here $xy$ LC is the only stable ordered phase, reached through a first order transition.
		The $xy$ and $x^2\! - \! y^2$ LC have two branches which develops as $R$ is lowered ({\bf 2a}), one stable (large $l$) and one unstable, as can be seen from the energy relation {\bf 2b}. 
		The saddle-point solution (green) represents an unstable branch $l_x \neq l_y >0$ that extrapolates between the $xy$ and $x^2\! \LCm \!y^2$ LC phase. In {\bf 2a} solid (dashed) green represents $l_x$ ($l_y$) connecting the $l_x=l_y$ solution with $l_x>0,l_y=0$. {\bf 2c} and {\bf 2d} shows the development of the B$_{1g}$ nematic order, $N_{x^2\! \LCm \!y^2}$, and the renormalized static susceptibility $r'$.
		{\bf 3):} Developing of order for $\alpha = 1.1$. Here, both $xy$ and $x^2\! - \!y^2$ LC are possible stable phases. At $R=1.02$ the system goes through a first-order transition to a state with finite $l$, {\bf 3a}, as well as finite nematic order, {\bf 3c}. At $R=0.36$ the $xy$ and $x^2\! - \!y^2$ LC becomes degenerate {\bf 3b} and another first-order transition to the $xy$ LC phase occurs {\bf 3a}, where the {\it primary} B$_{1g}$ nematic order is lost {\bf 3c}. The saddle-point solution again extrapolates between the $x^2\! - \!y^2$ and $xy$ LC state and near transition $R=0.36$ all three states are near degenerate {\bf 3b}. {\bf 3d} shows the renormalized static susceptibility $r'$.
	}
\end{figure*}

As a representative case of $0.5<\beta<1, \beta<\alpha$, the phase diagram and the evolution of the order parameters is presented in Figure \ref{figure_multi_beta_075} for $\beta=0.75$, $\alpha=0.7$ and $\alpha=1.1$. (For $\alpha=0.75$ the normal, $xy$ and $x^2\! - \!y^2$ LC phase all coexist. It is possible to show that this holds in general for $\alpha=\beta$.)

%\subsection{The $x^2\! - \!y^2$ and $xy$ LC transition \label{subsection_enhanced_sym}}
%
\begin{figure*}
    \includegraphics[width=1.0\textwidth]{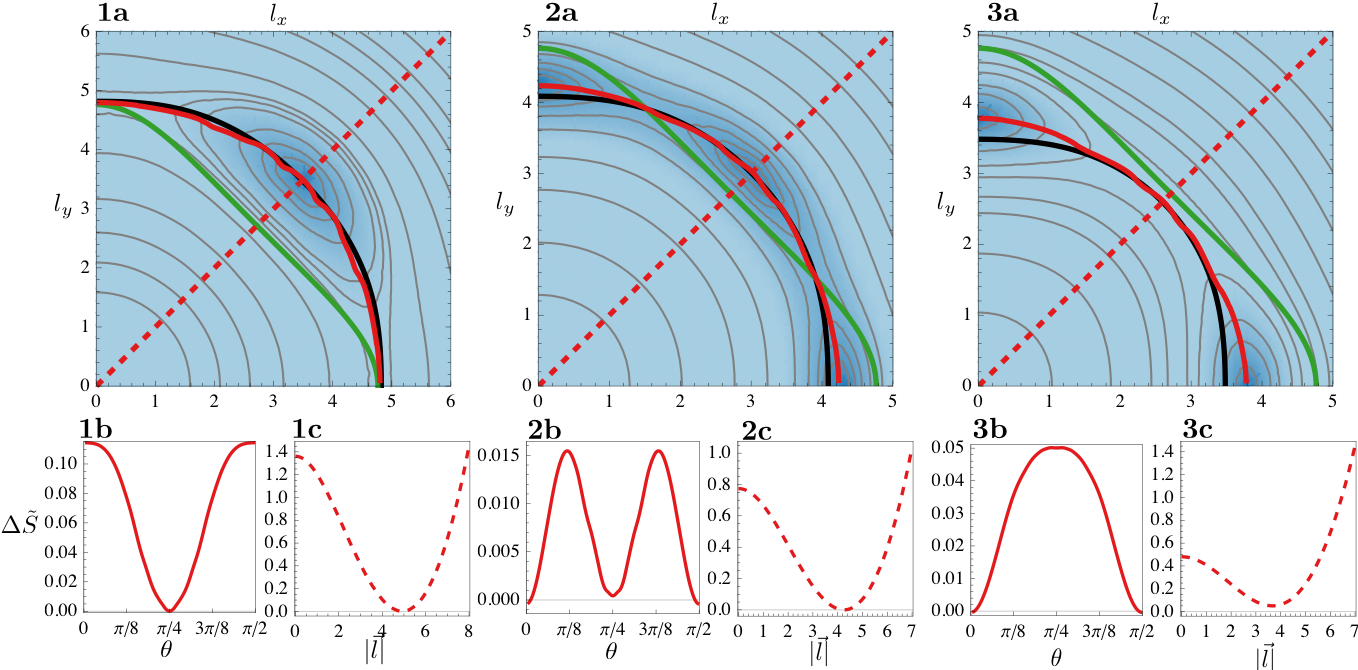} 
	\caption{\label{figure_near_deg}  \footnotesize Energy with respect to the ground state for arbitrary $\vec{l}$ from \eqref{energy_given_r} after solving for \eqref{mean:r} and \eqref{mean:phim} with $\beta=0.75, \alpha=1.1$ and $R=0.10;0.36;0.55$. {\bf a} equipotential contours of energy in gray, with dark shading indicating low energy. Solutions to \eqref{conditions:r} and \eqref{conditions:N} in 
	black and green respectively ($l_x=l_y$ solutions are removed from the latter). {\bf b-c} energy along the valley (solid red line) and the radial direction (dashed red line).
	{\bf 1} $R=0.10$, below the transition between the $xy$ and $x^2 \! - \!y^2$ LC phase in Figure \ref{figure_multi_beta_075}. The green line 
	%($l_x \neq l_y$ solutions to \eqref{conditions:N}) 
	do not cross the black line, 
	%(solutions to \eqref{conditions:r})
	thus there exist no intermediate extreme along the valley. $x^2 \! - \! y^2$, $xy$ LC are unstable and stable respectively, see {\bf 1b}. 
	%This is explicitly shown in {\bf 1b} where $xy$ LC corresponds to the minimum $\theta=\pi/4$, and $x^2 \! - \! y^2$ LC the maximum at $\theta=0,\pi/2$. 
	{\bf 1c} energy in the radial direction.
	{\bf 2a-c} Same as in {\bf 1a-c} for $R=0.36$ near the transition. {\bf 2a} intersection between the green and black curve, corresponding to intermediate extrema 
	%(the barrier-solution $l_x \neq l_y$) 
	along the valley direction. {\bf 2b} both $xy$ and $x^2 \! - \!y^2$ LC correspond to local minima. 
	%The energy landscape is essentially flat in the valley direction, compared to the radial, as can be seen by comparing to 
	{\bf 2c} energy in the radial direction.
	{\bf 3a-c} same as in {\bf 2a-c} for $R=0.55$ above the transition where $xy$ LC is stable and $x^2\! - \! y^2$ LC unstable, as can be seen in {\bf 3b}.
	}
\end{figure*}

The saddle-point solution ($l_x \neq l_y >0$) only has support for a finite range of $R \in (R_{\text{min}},R_{\text{max}})$, as indicated by the green regions in Figure \ref{figure_multi_beta_075}.3a-d, and forms closed paths in the $(l_x,l_y)$-plane (see also Figure \ref{1D_condR_condN_alpha_07_beta_075}a in Appendix \ref{sec_xy_LC}). As $R$ is tuned from $R_{\text{max}}$ to $R_{\text{min}}$, by lowering the temperature, $\vec{l}$ twists from $l_x=l_y$ with $N_{x^2\! \LCm \!y^2}=0$ to $l_x>0, l_y=0$ with $N_{x^2\! \LCm \!y^2}>0$. 

To explore this transition further we consider the absolute energy in terms of $\vec{l}$ \footnote{Care must be taken when expressing the energy solely in terms of $l_x,l_y$, see end of Appendix \ref{appendix_energy}}.
%
%we numerically solve \eqref{mean:r} and \eqref{mean:phim} given an arbitrary $\vec{l}$ by rewriting \eqref{mean:r},\eqref{mean:phim} as first order differential equations for $r'(\vec{l})$ and $N_{x^2\! \LCm \!y^2}(\vec{l})$, where \eqref{eq_r_normal} and $N_{x^2\! \LCm \!y^2}=0$ are taken as boundary conditions at $\vec{l}=0$. 
%
%The absolute energy in terms of $\vec{l}$ is found by inserting the solutions into \eqref{energy_given_r}. 
%
The energy is presented in Figure \ref{figure_near_deg} for $\alpha=1.1, \beta=0.75$, just below ($R=0.10$) and above ($R=0.55$) the support for the saddle-point solution, as well as near the transition $R=0.36$.

The $xy$ and $x^2 \! - \!y^2$ LC solutions lie on a semi-circular shaped valley in the energy landscape. Because of periodicity, the number of maxima equals the number of minima. Thus, in order for the $xy$ and $x^2 \! - \!y^2$ LC to be simultaneously stable, two intermediate maxima have to be introduced along the valley (in each quadrant). These are the $l_x \neq l_y$ solutions, and their energy can be seen as the height of the barrier between the two (meta)stable solutions. Nevertheless, these are saddle-point solutions in the full energy landscape. As is seen both from Figure \ref{figure_multi_beta_075}.3b and Figure \ref{figure_near_deg}.2b, this barrier height is small compared to the energy scale in the radial direction. Thus, the solutions are easily excited along the valley-direction. 

The relative smallness of the stiffness in the valley direction should be understood as a result of the valley direction being a compact dimension whose length is tunable to zero. Alternatively, it follows from expanding around a rotational invariant point $\vec{l}=0$. This effect is perhaps most easily seen from comparing Figure \ref{figure_near_deg}.{\bf 1,3} with Figure \ref{figure_near_deg}.{\bf 2}. In the former, the dispersion is only about twice as soft in the valley direction, while in the latter, the inclusion of three additional stationary points forces the dispersion to become even flatter in the valley direction. In the limit $\vec{l}\rightarrow 0$, this feature is expected to become more pronounced. To confirm, we expand \eqref{conditions:r} and \eqref{conditions:N} for small $\vec{l}$, and find
\begin{equation}
\begin{split}
\frac{6(1-R)}{\alpha-1}&= l_x^2 + l_y^2 + \scr{O}(l^4) \\
\frac{30(1-\beta) }{\beta}&= l_x^2 + l_y^2 + \scr{O}(l^4) \, \\
\end{split}
\label{eq_expansion_condition}
\end{equation}
describing two concentric circles. This implies that at $R= \frac{5-4 \beta }{\beta }+ \frac{5 (\beta -1)}{\beta }\alpha$ all points around the valley will be arbitrarily close to a local maximum, thus the valley direction will be essentially flat. 
%(Also, no singular behaviour arise in this limit.) 
%
This expansion becomes exact for $\beta \rightarrow 1^{-}$, and according to Table \ref{table_preemtive_sol}, there will be a second-order phase-transition for $\alpha>\beta$, where the $xy$ and $x^2-y^2$ LC states are degenerate. Thus, the first-order transition gets tuned into second-order transition. As a corollary, we expect a small stiffness in the valley-direction whenever there is only a small region of support for the $x^2-y^2$ solutions.

\section{Soft nematic state \label{section_soft}}
We have seen how the vicinity of the first-order transition between $xy$ and $x^2\! - \!y^2$ LC gives rise to an arbitrarily flat energy landscape, associated with a rotation of the LC order. Thus, a small field would be able to pin the LC order in any direction, promoting a state with, in general, both B$_{1g}$ and B$_{2g}$ nematic order. For concreteness, we can consider a correction to the superconducting mass, like the one found in \eqref{eq_effective_superconductive_action}, due to the LC order
%
%\begin{equation}
%S=
%\begin{bmatrix}
%g_1N_{\text{B}_{1g}} & g_2N_{\text{B}_{2g}}\\
%g_2N_{\text{B}_{2g}}& -g_1N_{\text{B}_{1g}} \\
%\end{bmatrix} \, ,\\
%\end{equation}
%
\begin{equation}
S=
\begin{bmatrix}
g_1(l_x^2-l_y^2)/2 & g_2l_xl_y\\
g_2l_xl_y& -g_1(l_x^2-l_y^2)/2 \\
\end{bmatrix} \, .
\end{equation}
%
%with $N_{\text{B}_{1g}}\propto l_x^2 - l_y^2$, $N_{\text{B}_{2g}}\propto l_xl_y$. 
The principal direction of $S$ will in general not be aligned with the crystallographic axis, as well as being easily pinned in any direction. We will refer to this as a ``soft'' nematic state. (Note that $S$ does not constitute an $XY$ nematic order parameter unless $g_1=g_2$.) 

There are some evidence for such a soft nematic state for the cuprates. A detachment of the nematic director from the lattice is seen in transport measurements on LSCO films\cite{wu2017spontaneous}. A signature which is strongest near optimum doping. This state is also expected to be sensitive to quenched disorder, which might explain the decreasing nematic domain size in BSSCO\cite{fujita2014simultaneous}, approaching optimum doping. This would suggest that the underlying nematic state seen both in LSCO and BSSCO is of the same origin; an underlying $xy$ ME-PDW preempted by a $xy$ and/or $x^2 \! - \!y^2$ LC state in turn setting up a soft nematic state, of the type described here; in LSCO the nematicity may be aligned by an external symmetry breaking field, while in BSCCO it is pinned to impurities. We include this scenario in the inset of Figure \ref{fig_pahase_diagram}, where the vestigial nematic phase is divided in a high and low temperature phase with $x^2-y^2$ and $xy$ nematic order respectively, with a soft nematic state at the boundary of the two phases.

\subsection{Emergent overdamped Goldstone mode \label{subsection_goldstone}}

The approximate rotational symmetry for the LC order near the first-order transition implies the existence of low-lying collective excitations. In the limit of an exact rotational symmetry this would corresponds to a Goldstone mode. To explore the the signatures of the soft nematic state near the transition we here consider the the spectral function of this ``emergent'' Goldstone mode.

The collective modes of the $xy$ and $x^2 \! - \!y^2$ LC states involves not only the LC order but couples all fields in the complete $\psi,N_{x^2\! \LCm \!y^2},\vec{l}$ space; as $\vec{l}$ change, $N_{x^2\! \LCm \!y^2}$ and $\psi$ changes accordingly. Thus, in general, a collective mode is associated with variations in the combined space of $\Phi=[\psi,N_{x^2\! \LCm \!y^2},l_x,l_y]$. 
In \eqref{eq_effective_action}, we neglected the time-dependence of the (bosonic) fields, corresponding to a high-temperature (classical) limit, where only the first bosonic Matsubara frequency is kept. Now we reintroduce the Matsubara frequencies to find the excitation spectra by analytic continuation to real frequencies. 

%Since fields in their Matsubara representation $f(i\omega_n)=\int_{0}^{1/T} \id \tau \, e^{-i\omega_n \tau} f(\tau)$ comes with an additional unit of energy, we extract factor of $1/T$ from the coefficients by writing $r=\breve{r}/T$.
%, \kappa = \breve{\kappa}/T, u = u/T, \gamma = \breve{\gamma}/T$. 
%\com{$\breve{r} \check{r} \hat{r}$}
%
The PDW part of the effective action \eqref{eq_effective_action} (for the A sector) can be written as
\begin{equation}
\begin{split}
& S_{\text{eff}}(\{ \Delta_{\ve{Q}} \},\psi,N_{x^2\! \LCm \!y^2},\vec{l})= \\
&-\frac{1}{T}\int \limits_{\mu} \frac{|\psi_{\mu}|^2}{2u_0} +\frac{|N_{x^2\! \LCm \!y^2,\mu}|^2}{2u_1} + \frac{|\vec{l}_{\mu}|^2}{2u_2}  
+\int \limits_{\mu,\mu'} \vec{\Delta}^{\dagger}_{\mu} \overline{\G}^{-1}_{\mu;\mu'} \vec{\Delta}_{\mu'} \\
\end{split}
\label{eq_effective_action_appendix_1}
\end{equation}
where we included an integral $T \int_0^{1/T} \id \tau$ and fields in the Matsubara representation
$f(i\omega_n)=\int_{0}^{1/T} \id \tau \, e^{-i\omega_n \tau} f(\tau)$, with the bosonic Matsubara frequencies $\omega_n=2\pi n T$. We introduced the short-hand notation $\mu = i\omega, \vec{k}$ and $\vec{\Delta}=[\Delta_{\QAx} \Delta_{-\QAx} \Delta_{\QAy} \Delta_{-\QAy}]$,  $\int_{\tau}=\int_{0}^{1/T} \id \tau$ and $\int \limits_{\mu}= T \sum \limits_{i \omega_n } \int \limits_{\vec{q}}$. We find the kernel $\G$ as
\begin{equation}
\begin{split}
\G^{-1}_{11,\mu;\mu'} &=(-i\omega+T\chi^{-1}_{x}(\vec{k}))\delta_{\mu,\mu'} \\ &+\psi_{\mu-\mu'}+ N_{x^2\! \LCm \!y^2,\mu-\mu'} + l_{x,\mu-\mu'}  \\
\G^{-1}_{22,\mu;\mu'} &=(-i\omega+T\chi^{-1}_{x}(\vec{k}))\delta_{\mu,\mu'}\\ &+\psi_{\mu-\mu'} + N_{x^2\! \LCm \!y^2,\mu-\mu'}- l_{x,\mu-\mu'}  \\
\G^{-1}_{33,\mu;\mu'} &=(-i\omega+T\chi^{-1}_{y}(\vec{k}))\delta_{\mu,\mu'}\\ &+\psi_{\mu-\mu'}  - N_{x^2\! \LCm \!y^2,\mu-\mu'} + l_{y,\mu-\mu'}  \\
\G^{-1}_{44,\mu;\mu'}&=(-i\omega+T\chi^{-1}_{y}(\vec{k}))\delta_{\mu,\mu'}\\ &+\psi_{\mu-\mu'} - N_{x^2\! \LCm \!y^2,\mu-\mu'} - l_{y,\mu-\mu'}  \\
\end{split}
\label{eq_kernel}
\end{equation}
where $\chi_{x,y}^{-1}=r+\kappa_1(k_x^2+k_y^2)  \pm \kappa_2(k_x^2-k_y^2)$, and $\delta_{\mu,\mu'}= \delta(\vec{k}-\vec{k}')  \delta_{i\omega,i\omega'}/T$. We assume that the PDW fields are coherently propagating, and not damped. (This would be the case if PDW arose from a strong-coupled BEC scenario\cite{waardh2018suppression}.)
The effective action is found in terms of $\Phi$ by integrating over the PDW fields. 
We proceed by expanding around the uniform mean-field solution $\frac{\delta S_{\text{eff}}}{\delta \Phi }=0$, given by $\Phi_0$, 
\begin{equation}
     \Phi(i\nu_n,\vec{q})=\Phi_0 \delta_{n,0}\delta(\vec{q}) + \delta \Phi(i\nu_n,\vec{q}) \, .
\end{equation}
Expanding the action to second order in $\delta \Phi$
\begin{equation}
\begin{split}
&S_{\text{eff}}(\psi,N_{x^2\! \LCm \!y^2},\vec{l}) \approx  \\
& S^{(0)}_{\text{eff}}(\psi_0,N_{x^2\! \LCm \!y^2,0},\vec{l}_0) + S^{(2)}_{\text{eff}}(\psi,N_{x^2\! \LCm \!y^2},\vec{l}) \, .
\end{split}
\label{eq_app_effective_decomp}
\end{equation}
In the high-temperature limit, keeping only the first Matsubara term $n=0$,  $S^{(0)}_{\text{eff}}(\psi,N_{x^2\! \LCm \!y^2},\vec{l})$ is given by \eqref{eq_S_mean_field} (with $\Delta_0$ reinserted). The correction can be written
\begin{equation}
\begin{split}
S_{\text{eff}}^{(2)}=& \int \limits_{\mu} \frac{1}{2}\delta \Phi_{i}(\mu) \scr{L}^{-1}_{ij}(\mu)\delta \Phi_{j}(\mu) \\
\scr{L}^{-1}_{ij}(i\nu_n,\vec{q})&=\frac{-\delta_{ij}}{Tu_{i}}-\ov{\Sigma}_{ij}(i\nu_n,\vec{q}) \, . 
\end{split}
\end{equation}
Here $\scr{L}$ is the propagator of fluctuations of $\Phi$, with $u_{i}=[u_0,u_1,u_2,u_2]$. The self-energy term is given by
\begin{equation}
\begin{split}
&\ov{\Sigma}=\\
&\left(
\begin{array}{cccc}
	\Sigma_{1}\LCp\Sigma_{2}\LCp\Sigma_{3}\LCp\Sigma_{4}& \Sigma_{1}\LCp\Sigma_{2}\LCm\Sigma_{3}\LCm\Sigma_{4} & \Sigma_{1}\LCm\Sigma_{2} & \Sigma_{3}\LCm\Sigma_{4} \\
	\Sigma_{1}\LCp\Sigma_{2}\LCm\Sigma_{3}\LCm\Sigma_{4} & \Sigma_{1}\LCp\Sigma_{2}\LCp\Sigma_{3}\LCp\Sigma_{4} & \Sigma_{1}\LCm\Sigma_{2} & \Sigma_{4}\LCm\Sigma_{3} \\
	\Sigma_{1}\LCm\Sigma_{2} & \Sigma_{1}\LCm\Sigma_{2} & \Sigma_{1}\LCp\Sigma_{2} & 0 \\
	\Sigma_{3}\LCm\Sigma_{4} & \Sigma_{4}\LCm\Sigma_{3} & 0 & \Sigma_{3}\LCp\Sigma_{4} \\
\end{array}
\right)
\end{split}
\label{eq_self_energy}
\end{equation}
where
\begin{equation}
\begin{split}
&\Sigma_{i}(i\nu_n,\vec{q})  =  \int \limits_{i \omega,\vec{k}} G_{i}(i\omega + i \nu,\vec{k}+\vec{q})G_{i}(i\omega ,\vec{k}) \, ,\\
\end{split}
\end{equation}
%
%where $G_1(i \omega, \vec{k})= -i\omega/T + \chi_{x}(\vec{k}) +$
%
and $G_{i}^{-1}(i\omega,\vec{k})=-i\omega +T\chi^{-1}_{\ve{Q}(i)}(\vec{k})$ where $\ve{Q}(i)=[\ve{Q}_1,-\ve{Q}_1,\ve{Q}_2,-\ve{Q}_2]$. The propagator $\scr{L}_{ij}(i\nu_n,\vec{q})$ is in general off-diagonal. However, when approaching the $x^2-y^2$ to $xy$ LC transition from the $x^2 \! - \!y^2$ LC state, with $l_x=l_0,l_y=0$ and $N_{x^2\! \LCm \!y^2} \neq 0$, $\overline{\Sigma}_{ij}(i\nu_n,\vec{q})$ is block diagonal,
%Inserting these conditions into \eqref{eq_self_energy} yields 
$\overline{\Sigma}_{ij}=\overline{\Sigma}_{ab} \oplus\overline{\Sigma}_{44}, a,b=1,2,3$. In this case the valley-direction lies solely along $l_y$, with all other fields stationary.

The static, zero-frequency part of the propagator $\scr{L}^{-1}_{ij}(0,0)$ is associated with the stiffness to uniform deformation. Here $\scr{L}^{-1}_{44}(0,0)$ can be related with the stiffness along the valley direction, which we set to zero and identify it with the transverse propagator for a nematic director along the $x$-axis, $\scr{L}^{-1}_{44}=\scr{L}^{-1}_{\perp,x}$. After analytic continuation of
%\scr{L}^{-1}_{44}(0,0)\propto \rho_{\perp}$
%
\begin{equation}
%\begin{split}
\overline{\Sigma}_{44}(i\nu_n,\vec{q})  =  2 \int \limits_{i \omega,\vec{k}} G_{4}(i\omega_n + i \nu_n,\vec{k}+\vec{q})G_{4}(i\omega_n, \vec{k}) \, .
\end{equation}
we can obtain the retarded transverse propagator. In the high-temperature limit, after expanding in $\vec{q}$ and $\nu/q$ (assuming $r\gg N_{x^2\! \LCm \!y^2}, \vec{l}$, i.e. far from the PDW transition) we find
\begin{equation}
\begin{split}
&\scr{L}^{-1}_{\perp,x}(\nu,\vec{q}) = \eta \tilde{q}^2 -i\eta' s+ \scr{O}(s \tilde{q}) + \scr{O}(s^3) \\
%&+  i\eta'_{1} \frac{\nu\tilde{q}}{r_a^{5/2}}   +i\eta'_{2} \frac{s_a^3}{r_a^{5/2}}  + \eta_{1}\frac{s_a \tilde{q}}{r_a^2}  -  \eta_{2}\frac{s_a^2\tilde{q}^2}{r_a^3}
\end{split}
\end{equation}
where $s=\nu/\tilde{q}$,
%$\eta= \frac{1}{12\pi \beta \overline{\kappa}} \frac{1}{(r-N_{x^2\! \LCm \!y^2})^2}$, $\eta' = \frac{1}{8\beta\overline{\kappa}} \frac{1}{(r-N_{x^2\! \LCm \!y^2})^{3/2}}$, $\tilde{q}=q\sqrt{\kappa_1 - \kappa_2 \cos(2\varphi)}$
$\eta= \frac{1}{12\pi T\overline{\kappa}} \frac{1}{r^2}$, $\eta' = \frac{1}{8T^{2}\overline{\kappa}} \frac{1}{r^{3/2}}$, $\tilde{q}=q\sqrt{\kappa_1 - \kappa_2 \cos(2\varphi)}$ where $\varphi$ is the angle of $\vec{q}$ to the $x$-axis.
\begin{figure}[h!]
	\centering	
	\includegraphics[width=0.4\textwidth]{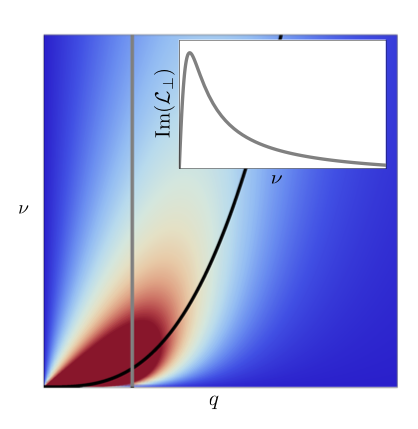} 
	\caption{\label{fig_spectral}  \footnotesize Spectral density of the Goldstone mode at the enhanced symmetry point $\text{Im}(\scr{L}_{\perp}(\nu,\vec{q}))$. The black line shows the peak of the spectral density at fixed $\vec{q}$. The inset shows the frequency dependence of the spectral density along the gray line. }
\end{figure}
In Figure \ref{fig_spectral}, the spectral density $\text{Im}(\scr{L}_{\perp})$ is shown, and we see an overdamped bosonic mode, with $\nu \propto i q^3$. %, due to Landau-damping.

These results are reminiscent of the results for a nematic Fermi fluid\cite{oganesyan2001quantum}, where an overdamped Goldstone mode is found within the broken symmetry phase. (The reason has to do with the non-commuting property of the broken symmetry and translation\cite{watanabe2014criterion}.)
In a fermionic system, an overdamped bosonic mode coupling to the fermions usually leads to the destruction of the Fermi liquid. This is of special interest in cuprates as a possible origin of the strange metal phase\cite{keimer2015quantum,varma1989phenomenology}.

A Goldstone mode associated with spatial rotation is not expected in a crystalline system since there is no rotational symmetry. However, as seen in this work, a competition between a vestigial $xy$ and a $x^2 \! - \!y^2$ LC phase leads to a very small gap and an emerging symmetry. It is intriguing to note that this phenomenology again points towards an emergent rotational symmetry near the overdoped critical point (possibly a soft nematic state), over which the strange metal phase is located.
Indeed, at this level, these results are only speculative. First of all, in the model considered here, the bosonic mode does not couple directly to fermions, but to PDW fluctuations. The fate of the PDW fields, and the underlying fermions are left for future work. %Here we instead report the general phenomenology, possibly emerging in other, symmetry related systems. 

\section{Summary and Outlook}

In this paper, we study vestigial orders of a PDW state with pair-momenta that are aligned with the high symmetry directions of a tetragonal crystal, focusing on phases that only break the point-group symmetry. Of particular interest is the influence of vestigial nematic order on a homogenous single component superconductor, giving rise to an anisotropic superfluid stiffness. We stress that if the nematicity arises from a PDW state, i.e., from finite momentum fluctuations of the superconducting field itself, the superconductor can become highly susceptible to this nematic distortion due to the natural proximity of a Lifshitz point with vanishing superfluid stiffness. Crucially, even a nominally weak nematic field, as observed through anisotropy of the normal electron response or the superconducting gap, may give a large relative renormalization of a small stiffness. 
%, while 
We argue that this may explain why the observed anisotropy in transport measurements on LSCO \cite{wu2017spontaneous} can be ascribed to highly anisotropic superconducting fluctuations coexisting with an essentially isotropic normal conductivity \cite{wardh2019inprep}. Probing vortex dynamics, also expected to be sensitive to stiffness anisotropy,  near $T_c$, could be a fruitful pursuit to investigate this unusual manifestation of nematicity further.

In the later part of the paper we focus on vestigial orders of a magnetoelectric (ME) PDW, containing both nematic order and loop-current type order.%, suggested to account for time-reversal breaking observed in the pseudogap phase. 
We have shown that a preemptive transition into a vestigial phase of an ME-PDW with $B_{1g}$ ($x^2-y^2$) nematic order can split into a high and low-temperature phase, that correspond to distinct $B_{1g}$ and $B_{2g}$ ($xy$) phases. This feature is not specific to PDW, but is expected for any other field transforming in the $A_{1g} \oplus B_{1g} \oplus E_u$ (or $A_{1g} \oplus B_{2g} \oplus E_u$) representation.
%Regarding an underlying ME-PDW as potential cuprate pseudogap state, it can explain both the occurrence of $xy$ and $x^2-y^2$ nematic order, as well as a mixture of the two. In fact, 
Near the transition between the high and low-temperature phases, the nematic order will be soft and easy to pin in either direction, yielding an approximate rotational symmetry, with possible relevance to observations of nematic order in LSCO and BSCCO.  
%Possible experimental support in the cuprate system was discussed.
%
%
Also, as a start for further investigation, the emergence of an overdamped Goldstone mode due to this approximate rotational symmetry may have interesting implications for the single-particle properties of electrons coupling to this mode.  
%could possibly be connected to non-Fermi liquid behavior. 

In conclusion, the results lend support and warrant further investigation into the proposal of pair-density wave order as the underlying source of the abundance of broken symmetries and exotic phenomenology seen in the cuprate superconductors.

Note: After the initial posting of this work an experimental study of overdoped Bi$_2$Sr$_2$CaCu$_2$O$_{8+x}$ using scanning Josephson tunneling microscopy has presented evidence for a nematic state with short range PDW order, interpreted as a disorder-pinned realization of a state with vestigial nematic PDW order\cite{chen2022identification}. %+Zink

\section{Acknowledgements}
We thank I. Bo{\v{z}}ovi{\'c} for valuable discussions.

% ========================= BIBLIOGRAPHY ====================

\bibliography{bib_archive_master.bib}

\appendix

%\section{Solutions to the Vestigial mean-field equations with LC order}
\section{$xy$ LC state and saddle-point solution, $l_x, l_y \neq 0$ \label{sec_xy_LC}}
\begin{figure}
	\centering
	\includegraphics[width=0.44\textwidth]{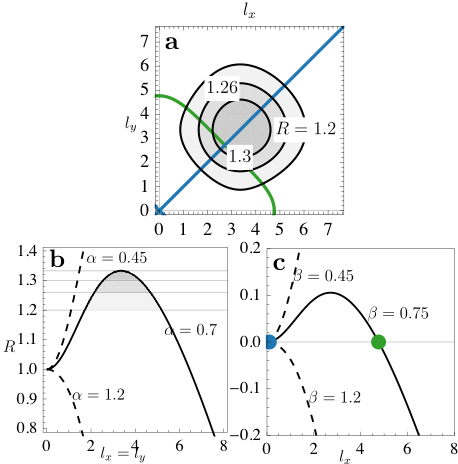} 
	\caption{\label{1D_condR_condN_alpha_07_beta_075}  \footnotesize 
		{\bf a):} Solutions of \eqref{conditions} in the $l_x, l_y$-plane. The black curves are solutions to \eqref{conditions:r} 
		$\alpha=0.7$ for $R=1.2,1.26,1.3,1.33$(peak $R$). The blue line corresponds to the $xy$ LC  $l_x = l_y$ solutions to \eqref{conditions:N} for $\beta=0.75$, while the saddle-point solution, $l_x \neq l_y$, is shown in green. Simultaneous solutions to \eqref{conditions} are given by the intersection of the blue (or green) line with the black line.
		{\bf b):}  Right-hand side of \eqref{conditions:r} for $\alpha = 0.45 , 0.7, 1.2$ along the $\vec{l}$ diagonal $l_x=l_y$, for which \eqref{conditions:N} is trivially fulfilled. For $\alpha<0.5$ ustable solutions exists for $R>1$. For $\alpha>1$ there are stable solutions for $R<1$. When $0.5<\alpha<1$ solutions occur for a finite $l$, corresponding to a first-order transition. %Intersections for $R=1.2,1.26,1.3, 1.333$(peak) is shown for $\alpha=0.7$. 
		{\bf c):} Right-hand side of \eqref{conditions:N} for $\beta= 0.45 , 0.75, 1.2$ plotted along the axial direction, $l_x>0, l_y=0$, as the condition is trivially zero for $l_x = l_y$. The blue dot indicates the intersection with the trivial zero-solutions (blue lines in {\bf a}), while the green dot indicates $l_x  \neq l_y$ solutions (green lines in {\bf a}). $l_x  \neq l_y$ solutions only exist for $0.5<\beta<1$. 
	}
\end{figure}

Let us start by assuming that both $l_{x}$ and $l_y$ are nonzero. All equations \eqref{mean} respects the symmetry $l_{x,y} \rightarrow - l_{x,y}$, thus we can focus on $l_{x,y}>0$. Non-trivial $l_x, l_y >0$ solutions of \eqref{mean:etax}\eqref{mean:etay} take the form 
\begin{equation}
r'\pm N_{x^2\! \LCm \!y^2} = - l_{x,y} \coth \left( \frac{2 \pi \ov{\kappa}}{u_2} l_{x,y}\right)
\label{eq_sol_general_eta}
\end{equation}
from which we see the need for $u_2 < 0$ to ensure $r' \pm N_{x^2\! \LCm \!y^2} - l_{x,y}>0$. The existence of primary B$_{1g}$ nematic order, $N_{x^2\! \LCm \!y^2} \neq 0$, is equivalent to $l_x \neq l_y$ as seen from \eqref{mean:phim} which ensures $N_{x^2\! \LCm \!y^2} \neq 0$ if $l_x \neq l_y$, while \eqref{mean:etax},\eqref{mean:etay} implies $N_{x^2\! \LCm \!y^2} = 0$ if $l_x = l_y$, unless $l_x = l_y=0$. 
Only considering the A sector we do not find any primary ($N_{xy}$) B$_{2g}$ nematic order for the $l_x=l_y$ solution. However, if both the A and B sector were stable, the subleading ($l_xl_y$) B$_{2g}$ nematic order would induce a finite primary B$_{2g}$ nematic order $N_{xy}$ (supported by sector B) through the coupling set by $v_{1}$ in \eqref{eq_full_action}. In Appendix \ref{section_coupling_axial_diagonal} we discuss the inclusion of both sectors, but for the present discussion this will be implicit.

Back to solving \eqref{mean}. Using \eqref{eq_sol_general_eta} to simplify  \eqref{mean:r} and \eqref{mean:phim} we receive two new (reduced) mean-field equations  
\begin{subequations}
	\label{conditions}
	\begin{align}
	\label{conditions:r}
	&R=\frac{\tilde{l}_{x} \coth \left( \tilde{l}_{x}  \right) + \tilde{l}_{y} \coth  \left( \tilde{l}_{y}  \right)}{2}  
	\\
	&	+\alpha \ln \left( \frac{\tilde{l}_{x} \tilde{l}_{y} }{\sinh \left( \tilde{l}_{x} \right)\sinh \left( \tilde{l}_{y} \right)}  \right) 
	\notag
	\\
	\label{conditions:N}
	&0=\frac{\tilde{l}_{x} \coth \left( \tilde{l}_{x}  \right) - \tilde{l}_{y} \coth  \left( \tilde{l}_{y}  \right) }{2} 
	\\
	&+ \beta\ln \left( \frac{\tilde{l}_{x}\sinh \left( \tilde{l}_{y} \right)}{ \tilde{l}_{y} \sinh \left( \tilde{l}_{x}  \right)} \right) 
		\notag
	\end{align}
\end{subequations}
where we introduced the normalization $\tilde{l}_{x,y} = \frac{2 \pi \ov{\kappa}l_{x,y}}{|u_2|}$, and $R  = \tilde{r}^R - 2\alpha \ln \left( \frac{|u_2|}{2 \pi \ov{\kappa}}  \right)$.
%
% as well as $\alpha =  \frac{u_0}{|u_2|}, \beta =  \frac{u_1}{|u_2|}$. 
Note that equation \eqref{eq_sol_general_eta} is not valid in the limit $l_{x,y} \rightarrow 0$, since it implies $r' \pm N_{x^2\! \LCm \!y^2} = - \frac{u_2}{2 \pi \ov{\kappa}}$, whereas \eqref{mean:etax} and \eqref{mean:etay} do not put any constraints on $r' \pm N_{x^2\! \LCm \!y^2}$. Thus $l_{x,y} = 0$ must be considered independently.

We find $l_x$ and $l_y$ by simultaneously solving \eqref{conditions:r} and  \eqref{conditions:N}, which we can interpret graphically as in Figure \ref{1D_condR_condN_alpha_07_beta_075}. Equation \eqref{conditions:N} is always solved for $l_x = l_y,N_{x^2\! \LCm \!y^2}=0$, corresponding to (meta)stable $xy$ solutions. Equation \eqref{conditions:r} determines the evolution of $\vec{l}$, as $R$ is changed. For $\alpha>1$ there is an onset of stable solutions at $R=1$ and $\vec{l}=0$, corresponding to a second-order phase-transition. For $0.5 < \alpha < 1$ a locally stable solutions occur at a finite $\vec{l}$ for some $R>1$, implying a first-order phase-transition. %These results are included in Table \ref{table_preemtive_sol}.

Equation \eqref{conditions:N} also admits solutions with $l_x \neq l_y, N_{x^2\! \LCm \!y^2} \neq 0$ for $0.5 < \beta<1$, as can be seen from Figure \ref{1D_condR_condN_alpha_07_beta_075}. These solutions, only supported for a finite range in $R$, will evolve along a curved path in the $l_x,l_y$-plane, with a corresponding change in $N_{x^2\! \LCm \!y^2}$ as $R$ changes. These solutions are unstable; however, as will be clarified, their existence indicates that a $xy$ and a $x^2-y^2$ LC state are simultaneously stable. 
%The energy of the $l_x \neq l_y >0$ solution corresponds to the height of the barrier between the two (meta)stable solutions, and constitute a saddle-point in the energy landscape. 
We will refer to this solution as the saddle-point solution since it constitutes a saddle-point in the energy landscape. Indeed, in the Section \ref{section_comp} we will see how $0.5 < \beta<1$ admits a first-order transition between the $xy$ and $x^2 \! - \!y^2$ LC states, with the possibility of a superheated and supercooled phase.

\section{\texorpdfstring{$x^2\! \LCm \!y^2$}{} LC state, \texorpdfstring{$l_{x,y}\neq 0, l_{y,x} =0$}{TEXT} \label{sec_xx+yy_LC}}
Now we assume the stability of the $x^2\! - \! y^2$ LC solutions with only one finite LC order component, say $l_x > 0$. Equations \eqref{mean:etay} puts no constraint on $r'- N_{x^2\! \LCm \!y^2}$, while \eqref{mean:etax} still implies $r' + N_{x^2\! \LCm \!y^2} = - l_{x} \coth \left( \frac{2 \pi \ov{\kappa}}{u_2}  l_{x}\right)$. For physical solutions we must require $u_2<0$ as well as $r' - N_{x^2\! \LCm \!y^2}>0$. In this case, we can directly solve for $r'$ and $N_{x^2\! \LCm \!y^2}$ using \eqref{mean:r} and \eqref{mean:phim}
\begin{subequations}
	\label{sol2}
	\begin{align}
	\label{sol2:r}
	%\ti{r}'&= \ti{l}_x\coth \left(\ti{l}_x \right) - \ti{N_{x^2\! \LCm \!y^2}}
	&R= \\
	\notag 
	&2 \alpha \ln \left(\frac{\ti{l}_x}{\sinh(\ti{l}_x)} \right) + \ti{l}_x \coth(\ti{l}_x) + (\alpha/\beta - 1 )\ti{N}_{x^2\! \LCm \!y^2}
	\\
	\label{sol2:N}
	&\ti{N}_{x^2\! \LCm \!y^2} = \\ \notag 
	&\frac{\ti{l}_x}{2}\coth \left(\ti{l}_x \right) - \beta W \left( \frac{ \ti{l}_x}{2 \beta} \frac{\exp \left[ \frac{\ti{l}_x}{2 \beta}\coth \left( \ti{l}_x \right) \right] }{\sinh \left( \ti{l}_x \right) } \right)
	\end{align}
\end{subequations}
where $W(x)$ denotes the product-logarithm. 
%We find $r'$ as $\ti{r}'= \ti{l}_x\coth \left(\ti{l}_x \right) - \ti{N}_{x^2\! \LCm \!y^2}$. 
(For $l_y \neq 0, l_x =0$ take $l_x \rightarrow l_y$ and $N_{x^2\! \LCm \!y^2} \rightarrow -N_{x^2\! \LCm \!y^2}$.) 
%
% including overall condition \alpha>1/2 for \beta>1/2 and \alpha>1-\beta for \beta<1/2
With $R \propto T$, non-trivial solutions evolves as $R$ is lowered and \eqref{sol2:r} starts admitting solutions. For $\beta>0.5$ and $\alpha> \frac{2+\beta}{4 \beta -1}$ 
% $\alpha> \text{max}(\frac{2+\beta}{4 \beta -1},1/2)$ 
% including overall condition \alpha>1/2 for \beta>1/2 and \alpha>1-\beta for \beta<1/2
there is an onset of solutions at $R=1,l_x=0$, corresponding to a second-order phase-transition. For $0.5<\alpha <\frac{2+\beta}{4 \beta -1}$ solutions occur at a finite $l_x$, corresponding to a first-order phase-transition. With similar arguments for $\beta<0.5$, we find the transitions included in Table \ref{table_preemtive_sol}.

\section{Vestigial mean-field energy \label{appendix_energy}}
% prev : Solving the mean-field equations - coexistence of loop-current and nematic order
%
The mean-field solutions only guarantee local stability, and we must compare the absolute energy of the different phases in order to find the ground state. Therefore, we need the normal-state solutions $N_{x^2\! \LCm \!y^2}=0, l_{x,y}=0$, as well, which are always admitted by \eqref{mean:phim}\eqref{mean:etax}\eqref{mean:etay}. Equation \eqref{mean:r} is readily solved by
\begin{equation}
\begin{split}
\ti{r}' =2\alpha W \left( \frac{e^{\frac{R}{2\alpha}}}{2\alpha} \right) \, ,
\end{split}
\label{eq_r_normal}
\end{equation}
requiring $r'>0$. 

The energy \eqref{eq_action_2D} has an explicit dependence on the cutoff $\Lambda$, which also renormalizes $r$. In the limit $\Lambda \rightarrow \infty$ the cutoff dependency can be absorbed in a constant energy term, given that the mean-field solution \eqref{mean:r} fulfilled 
%(see Appendix \ref{appendix_2D_energy_mean_field})
%
\begin{equation}
\begin{split}
&
\ti{S} = %-\sgn(u_2) 
\frac{\vec{\ti{l}}^2}{2} 
- \frac{\ti{l}_{x}}{2} \ln \frac{\ti{r}'+\ti{N}_{x^2\! \LCm \!y^2}+\ti{l}_x}{\ti{r}'+\ti{N}_{x^2\! \LCm \!y^2}-\ti{l}_x}  
-\frac{\ti{l}_{y}}{2} \ln \frac{\ti{r}'-\ti{N}_{x^2\! \LCm \!y^2}+\ti{l}_y}{\ti{r}'-\ti{N}_{x^2\! \LCm \!y^2}-\ti{l}_y}  
\\
&
-  \frac{\alpha}{8}  
\ln^2((\ti{r}'+\ti{N}_{x^2\! \LCm \!y^2})^2-\ti{l}_x^2)((\ti{r}'-\ti{N}_{x^2\! \LCm \!y^2})^2-\ti{l}_y^2)
+2 \ti{r}'
\\
& -\frac{\ti{r}'}{2}\ln((\ti{r}'+\ti{N}_{x^2\! \LCm \!y^2})^2-\ti{l}_x^2)((\ti{r}'-\ti{N}_{x^2\! \LCm \!y^2})^2-\ti{l}_y^2)
\\
& -\frac{\ti{N}_{x^2\! \LCm \!y^2}^2}{2 \beta} - \frac{\ti{N}_{x^2\! \LCm \!y^2}}{2} \ln \left( \frac{(\ti{r}'+\ti{N}_{x^2\! \LCm \!y^2})^2-\ti{l}_x^2}{(\ti{r}'-\ti{N}_{x^2\! \LCm \!y^2})^2-\ti{l}_y^2} \right)+ \text{constant} \, ,
\end{split}
\label{energy_given_r}
\end{equation}
where $\ti{S}=\frac{S_0}{A}\frac{4 \pi^2 \ov{\kappa}^2}{|u_2|}$. This energy was used to find the stable vestigial phases listed in Table \ref{table_preemtive_sol}. 

It is important to note that the energy in \ref{energy_given_r} only holds if \eqref{mean:r} is fulfilled. In order to present the energy as a function solely on $\vec{l}$, as in Figure \ref{figure_near_deg}, we numerically solve \eqref{mean:r} and \eqref{mean:phim} given an arbitrary $\vec{l}$ by rewriting \eqref{mean:r},\eqref{mean:phim} as first order differential equations. As boundary conditions \eqref{eq_r_normal} and $N_{x^2\! \LCm \!y^2}=0$ were used at $\vec{l}=0$. The absolute energy in terms of $\vec{l}$ was then found by inserting the solutions into \eqref{energy_given_r}.

\section{Additional solutions to the vestigial mean-field equations \label{appendix_additional_mean_field}\label{appendix_nematic} \label{appendix_negative_u1}}

\begin{figure}[h!]
	\centering
	\includegraphics[width=0.35\textwidth]{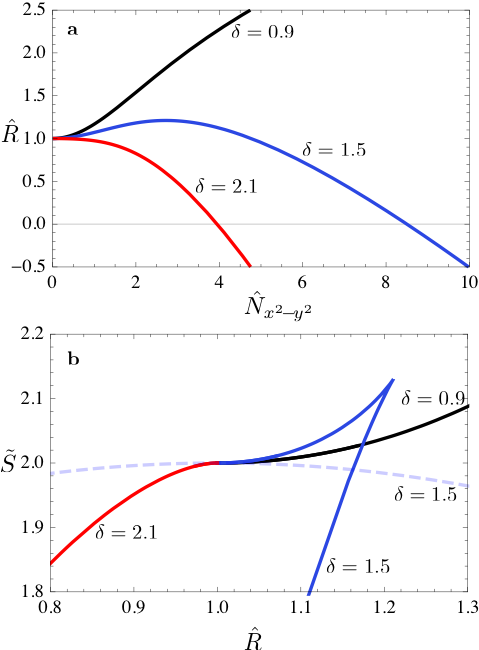} 
	\caption{\label{figure_pure_nematic} \footnotesize Evolution of the pure nematic phase for a range of $\delta=\alpha/|\beta|$. The local stability of the pure nematic phase is only dependent on the parameter $\delta$ while the absolute value of $\beta$ renormalizes the effective values of $R,N_{x^2\! \LCm \!y^2}$ and $S$, thus affecting the relative stability compared to other phases. 
	Here we used $\beta = -0.5$ for which $\hat{R}=R, \hat{N}_{x^2\! \LCm \!y^2}=\ti{N}_{x^2\! \LCm \!y^2}$. {\bf a)} Equation \eqref{eq_sol_N_no_eta_2_renormalized} plotted for a range of $\delta$. For $\delta<1$ there is no stable solution while for $\delta>2$ stable solutions occur for $R<1$ and the transition is second-order. For $1<\delta<2$ there is one stable and one unstable branch and the transition will be first-order. 
	}
\end{figure}

One set of solutions that were not considered in the main development is that of only primary nematic order without loop-current(LC) order $N_{x^2\! \LCm \!y^2} \neq 0, l_{x,y}=0$, the pure nematic phase. Equation \eqref{mean:etax},\eqref{mean:etay}  always admits the $l_{x,y}=0$ solution, and puts no constraint on $N_{x^2\! \LCm \!y^2}$ and $r'$. Non-trivial $N_{x^2\! \LCm \!y^2} \neq 0$ solutions to \eqref{mean:phim} take the form
\begin{equation}
r'= - N_{x^2\! \LCm \!y^2}\coth \left( \frac{\pi \ov{\kappa}}{u_1}N_{x^2\! \LCm \!y^2} \right) \, .
\label{eq_sol_N_no_eta}
\end{equation}
From which we find the expected requirement that we need $u_1<0$, in order for $r'\pm N_{x^2\! \LCm \!y^2} >0$. Assuming $u_1<$ we can rewrite \eqref{eq_sol_N_no_eta} as
\begin{equation}
\hat{r}'= \hat{N}_{x^2\! \LCm \!y^2}\coth \left( \hat{N}_{x^2\! \LCm \!y^2} \right) \, ,
\label{eq_sol_N_no_eta_renormalized}
\end{equation}
which inserted in \eqref{mean:r} yields

\begin{equation}
\begin{split}
&\hat{R}= \hat{N}_{x^2\! \LCm \!y^2}\coth \left( \hat{N}_{x^2\! \LCm \!y^2} \right) + \frac{\alpha}{|\beta|} \ln \left( \frac{\hat{N}_{x^2\! \LCm \!y^2}}{\sinh ( \hat{N}_{x^2\! \LCm \!y^2} ) } \right) \, , 
\end{split}
\label{eq_sol_N_no_eta_2_renormalized}
\end{equation}
where $\hat{N}_{x^2\! \LCm \!y^2} = \frac{ \pi \ov{\kappa}N_{x^2\! \LCm \!y^2}}{|u_1|}= \frac{\tilde{N}_{x^2\! \LCm \!y^2}}{2|\beta|}$ and  $\hat{R}=\frac{R}{2|\beta|}-\frac{\alpha}{|\beta|}\ln(2 |\beta|)$.
Equation \eqref{eq_sol_N_no_eta_2_renormalized} admits similar solutions as the $xy$ LC case \eqref{conditions:r} (see Figure \ref{figure_pure_nematic}), where we introduced $\delta = \frac{\alpha}{|\beta|}=\frac{u_0}{|u_1|}$. For $\delta<1$ there is one unstable branch for $\hat{R}>1$ and none for $\hat{R}<1$. For $1<\delta<2$ there is an unstable branch for small $\hat{N}_{x^2\! \LCm \!y^2}$ and a stable one for bigger $\hat{N}_{x^2\! \LCm \!y^2}$, which leads to a first-order transition. For $\delta>2$ there is one stable branch for $\hat{R}<1$ and $\hat{N}_{x^2\! \LCm \!y^2}$ evolves continuously from zero, yielding a second-order transition.

So far $u_2$ has not entered the analysis and the $N_{x^2\! \LCm \!y^2} \neq 0, l_{x,y}=0$ solution is locally stable as long as $u_1<0$, regardless of $u_2$.
For $u_2>0$ the pure nematic phase is the only locally stable solution. For $u_2<0$, the $xy$ and $x^2\! - \!y^2$ LC phases are in general stable as well and we have to compare the the absolute energy \eqref{energy_given_r} of the different phases. 
Taking into account the general stability of \eqref{eq_full_action} $\alpha>1-\beta$ for $\beta<0.5$ we find that the pure nematic phase is stable for $\beta<-0.5$ and the $xy$ LC phase is stable for $-0.5<\beta<0$.

\section{Coupling between A and B sector \label{section_coupling_axial_diagonal}}
\begin{figure}[h!]
	\centering	
	\includegraphics[width=0.4\textwidth]{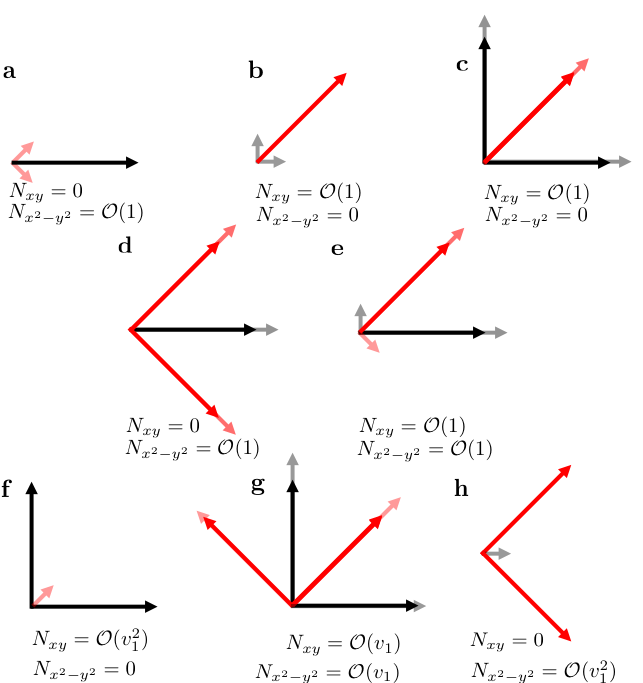} 
	\caption{\label{fig_perp_nematic}  \footnotesize Effect of couplings between the A and B sectors. Solid black (red) arrows corresponds to the unperturbed state in the A (B) sector $v_1=0$, whereas the shaded arrows corresponds to induced components (first-order in $v_1$). {\bf a-e} No additional primary nematic order. {\bf f-h} Additional primary nematic order is induced.}
\end{figure}

Bilinears of our model transforms in two distinct sectors, which we denote A and B. The stability of these two sectors, determined by the local minima of the dispersion \eqref{eq_dispersion}, are independent, and we can consider situations where either one or both of the sectors are present.

The structures of both sectors are identical, and the above analysis holds equally for both sectors if we take into account that the states refer to the principal axes of \newline \newline the A and B sector frame respectively, which are rotated 45$^{\circ}$ to each other. For concreteness, the $x^2-y^2$ LC and $xy$ LC phase of the A sector map to the $\ov{x}^2-\ov{y}^2=xy$ LC and $\ov{x}\ov{y}=x^2-y^2$ LC phase of the B sector.

If both sectors are present they will couple through fourth order terms, tuned by $v_{0,1}$ in \eqref{eq_full_action}. We will study this situation by including a weak interaction between the two sectors. Including non-zero couplings  $v_{0,1}$ in \eqref{eq_full_action} means that the matrix $M$ in the Hubbard-Stratonovich transformation \eqref{eq_HS}, will no longer be diagonal. Instead it will take the form 
\begin{equation}
M=
\begin{bmatrix}
M_{\sectA} & M_{\sectA\sectB} \\ 
M_{\sectA\sectB}^{T} & M_{\sectB} \\ 
\end{bmatrix}
\, , \,
M_{\sectA\sectB} = 
\begin{bmatrix}
v_0 & 0 & 0 & 0 \\
0 & 0 & 0 & 0 \\
0 & 0 & \frac{v_1}{\sqrt{2}} & -\frac{v_1}{\sqrt{2}}  \\
0 & 0 &\frac{v_1}{\sqrt{2}}   &  \frac{v_1}{\sqrt{2}}  \\
\end{bmatrix}
\end{equation}
and the auxiliary field vector $\Psi = [\psi_A, N_{x^2-y^2}, l_{\sectA, x},l_{\sectA, y},\psi_, N_{xy}, l_{\sectB ,\ov{x}},l_{\sectB ,\ov{y}}]$. Here $l_{\sectB ,\ov{x}},l_{\sectB ,\ov{y}}=\frac{l_{\sectB ,x}+l_{\sectB ,y}}{\sqrt{2}},\frac{l_{\sectB ,y}-l_{\sectB ,x}}{\sqrt{2}}$ is the B components along the ($xy$) diagonals. After integrating out the PDW field the effective action is given by (neglecting the superconducting field, which can be analogously introduced as before)
\begin{equation}
\begin{split}
&S_{\text{eff}}(\psi_{\sectA },N_{x^2-y^2},\vec{l}_{\sectA },\psi_{\sectB },N_{xy},\vec{l}_{\sectB }) =\\
&S_{\text{eff},\ti{A}}(\psi_{\sectA },N_{x^2-y^2},\vec{l}_{\sectA })+S_{\text{eff},\ti{B}}(\psi_{\sectB },N_{xy},\vec{l}_{\sectB }) \\
&- \left(  \frac{\psi_{\sectA } \psi_{\sectB } }{\ti{v}_0} + \frac{\vec{l}_{\sectA } \cdot\vec{l}_{\sectB }  }{\ti{v}_1} \right) \, .
\end{split}
\label{eq_something}
\end{equation}
The third term represents interaction between the A and B sectors, and the first two terms referring to the action of the A and B sectors respectively \eqref{eq_action_2D}. Due to the inversion of the off-diagonal matrix, $M$, the couplings become renormalized (indicated by the tilde in \eqref{eq_something}) $\ti{u}_{0,(\sectA\sectB)} = u_{0,(\sectA\sectB)}- \frac{v_0^2}{u_{0,(\sectA\sectB)}}$, $\ti{u}_{2,(\sectA\sectB)} = u_{2,(\sectA\sectB)}- \frac{v_1^2}{u_{2,(\sectA\sectB)}}$, $\ti{v}_0 = v_0- \frac{u_{0,A}u_{0,B}}{v_0}$ and $\ti{v}_{1}= v_1- \frac{u_{2,A}u_{2,B}}{v_{1}}$. 
The two sectors do not couple directly through the primary nematic order-parameters $N_{x^2-y^2}, N_{xy}$. The bilinear term in the LC orders of the two sectors implies mutual induction: A finite LC order in one sector will induce LC order in the other sector. This is evident from the (new) mean-field equations
\begin{subequations}
	\label{full_mean}
	\begin{align}
	\label{full_mean:psi}
	&\psi_{\sectA} = \psi_B \frac{v_0}{ u_{0,B}} +\frac{\ti{u}_{ 0,A}}{\pi\ov{\kappa}_{\sectA}} \ln(\ov{\kappa}_{\sectA}\Lambda_{\sectA}^2) \\
	\notag
	&-\frac{\ti{u}_{0,A}}{4\pi \ov{\kappa}_{\sectA}} \ln[(r'_{\sectA}+N_{x^2-y^2})^2-l_{\sectA x}^2][(r'_{\sectA}-N_{x^2-y^2})^2-l_{\sectA y}^2] \\
	\label{full_mean:phim}
	&N_{x^2\! - \! y^2}= - \frac{u_{1,A}}{4 \pi \ov{\kappa}_{\sectA}} 
	\ln \frac{(r'_{\sectA}+N_{x^2-y^2})^2-l_{\sectA x}^2}{(r'_{\sectA}-N_{x^2-y^2})^2-l_{\sectA y}^2}
	\\
	\label{full_mean:etaxy}
	&l_{\sectA,(x,y)}= \frac{ l_{ \ov{x},B}\mp l_{\ov{y},B}}{\sqrt{2}} \frac{v_2}{u_{2,A}} \\
	&-\frac{\ti{u}_{2,A} }{4 \pi \ov{\kappa}_{\sectA}}  
	\ln \frac{r'_{\sectA}\pm N_{x^2-y^2}+l_{\sectA,(x,y)}}{r'_{\sectA } \pm N_{x^2-y}-l_{\sectA,(x,y)}}  
	\\
	\notag
	\end{align}
\end{subequations}
(analogously for the B sector). Assuming a weak mixing $v_{0} \ll u_{0,(\sectA\sectB)}$, $v_1 \ll u_{2,(\sectA\sectB)}$ we can expand in orders of $v_{0,1}$. To first-order in $v_{0,1}$ we can solve the system by asserting $l=l^{(0)}+l^{(1)}$, $\psi=\psi^{(0)}+\psi^{(1)}$,  where $l^{(0)},\psi^{(0)}$ are solutions to the uncoupled case $v_{0,1}=0$. To first order 
\begin{widetext}
	\begin{subequations}
		\begin{align}
		&\psi^{(1)}_{\sectA} \left(1 + \frac{u_{0,\sectA}}{4\pi \ov{\kappa}_{\sectA}} \chi_{\sectA,\Sigma \ve{Q}} \right)= \psi^{(0)}_{\sectB} \frac{v_0}{ u_{0,\sectB}} %
		-\frac{u_{0,\sectA}}{4\pi \ov{\kappa}_{\sectA}}\Big[  
		\chi_{\sectA,\QAx}(0)( N_{x^2\!-\!y^2}^{(1)} + l_{\sectA,x}^{(1)} ) 
		+\chi_{\sectA,-\QAx}(0)(N_{x^2\!-\!y^2}^{(1)} - l_{\sectA,x}^{(1)} ) \\ 
		&-\chi_{\sectA,\QAy}(0)(N_{x^2\!-\!y^2}^{(1)} - l_{\sectA,y}^{(1)} )
		-\chi_{\sectA,-\QAy}(0)( N_{x^2\!-\!y^2}^{(1)} + l_{\sectA,y}^{(1)} ) \Big] 
		\notag \\
		&N_{x^2\!-\!y^2}^{(1)} \left(1 + \frac{u_{1,\sectA }}{4\pi \ov{\kappa}_{\sectA}}  \chi_{\sectA,\Sigma \ve{Q}} \right)= - \frac{u_{1,\sectA }}{4 \pi \ov{\kappa}_{\sectA}} 
		\Big[  
		\chi_{\sectA,\QAx}(0)(\psi_{\sectA}^{(1)} + l_{\sectA,x}^{(1)} ) 
		+\chi_{\sectA,-\QAx}(0)(\psi_{\sectA}^{(1)} - l_{\sectA,x}^{(1)} ) \\
		&
		-\chi_{\sectA,\QAy}(0)(\psi_{\sectA}^{(1)} + l_{\sectA,y}^{(1)} ) 
		-\chi_{\sectA,-\QAy}(0)(\psi_{\sectA}^{(1)} - l_{\sectA,y}^{(1)} ) \Big] 
		\notag \\
		&l^{(1)}_{\sectA x}(1+\frac{u_{2,\sectA } }{4 \pi \ov{\kappa}_{\sectA}}  \chi_{\sectA,\Sigma \ve{Q}_1} )=  \frac{ l^{(0)}_{\sectB \ov{x}}- l^{(0)}_{\sectB \ov{y}}}{\sqrt{2}} \frac{v_2}{u_{2,\sectA }}
		-\frac{u_{2,\sectA } }{4 \pi \ov{\kappa}_{\sectA}}   
		(\chi_{\sectA,\QAx}(0)-\chi_{\sectA,-\QAx}(0) ) (\psi_{\sectA}^{(1)} + N_{x^2\! - \! y^2}^{(1)} ) \\
		&l^{(1)}_{\sectA y}(1+\frac{u_{2,\sectA } }{4 \pi \ov{\kappa}_{\sectA}}  \chi_{\sectA,\Sigma \ve{Q}_2} )= \frac{ l^{(0)}_{\sectB \ov{x}}+ l^{(0)}_{\sectB \ov{y}}}{\sqrt{2}} \frac{v_2}{u_{2,\sectA }}
		-\frac{u_{2,\sectA } }{4 \pi \ov{\kappa}_{\sectA}}   
		(\chi_{\sectA,\QAy}(0)-\chi_{\sectA,-\QAy}(0) ) (\psi_{\sectA}^{(1)} - N_{x^2\! - \! y^2}^{(1)} )\\
		\notag
		\end{align}
	\end{subequations}
\end{widetext}
and similar for the B sector. Here we used the static susceptibilities of the unperturbed state
\eqref{eq_static_succeptibilities} and introduced $\chi_{\sectA,\Sigma\ve{Q}}= \chi_{\sectA,\QAx}(0) + \chi_{\sectA,-\QAx}(0) +\chi_{\sectA,\QAy}(0)+\chi_{\sectA,-\QAy}(0)$
$\chi_{\sectA,\Sigma\ve{Q}_{1,2}}= \chi_{\sectA,\ve{Q}_{1,2}}(0) + \chi_{\sectA,-\ve{Q}_{1,2}}(0)$. 
The correction to the vestigial mean-field solutions because of finite coupling between the sectors is illustrated in Figure \ref{fig_perp_nematic}, where black (red) arrows indicate the LC order in the A(B) sector. 
In Figure \ref{fig_perp_nematic}a the A sector is ordered in its $x^2-y^2$ LC state ($l_{\sectA,x}>0,l_{\sectA,y}=0$), while the B sector is not ordered to zeroth order in $v_1$. By turning on the coupling, an LC order in the B sector is induced $l_{\sectB,x}=l_{\sectB,y} = \scr{O}(v_1)$. This state, being symmetric for reflections in the $x$-axis, already has a finite expectation value for the primary nematic field, $N_{x^2-y^2}$, in the unperturbed case (analogous case for the B sector is shown in Figure \ref{fig_perp_nematic}b). Similar cases is shown Figure \ref{fig_perp_nematic}c-e where no additional primary nematic order is induced, since it is present already for $v_1=0$. 

In contrast, for the case noted above, an $xy$ LC phase in the A sector, which only have subleading $xy$ nematic order, $l_{\sectA,x}l_{\sectA,y}$, the coupling will induce a primary $xy$ nematic $N_{xy}$ (stemming from the B sector). This is depicted in Figure \ref{fig_perp_nematic}f (and similarly in g,h).

\end{document}